\documentclass[11pt,showpacs,amsmath,nofootinbib,superscriptaddress]{revtex4}
\usepackage{amssymb}
\usepackage[hypertex]{hyperref}
\usepackage{graphicx}
\usepackage{epstopdf}
\usepackage{bm}
\usepackage{epsfig}
\usepackage{graphics}
\usepackage{xspace}
\usepackage{amsfonts}

\def\Slash#1{{#1\!\!\!\slash}}

\def\nslash{n\!\!\!\slash}
\def\bnslash{\bar n\!\!\!\slash}

\def\OMIT#1{}

\newcommand{\nn}{\nonumber}

\newcommand{\bn}{{\bar n}}
\newcommand{\bea}{\begin{eqnarray}}
\newcommand{\eea}{\end{eqnarray}}

\newcommand{\mcdot}{\!\cdot\!}

\newcommand{\Gcusp}{\Gamma_{\rm cusp}}
\newcommand{\half}{\frac{1}{2}}

\newcommand{\muj}{\mu_j}

\newcommand{\cO}{{\cal O}}

\newcommand{\la}{\langle}
\newcommand{\ra}{\rangle}

\begin{document}


\title{Resummation prediction on top quark transverse momentum distribution at large $p_T$}

\vspace*{1cm}

\author{Jian Wang}
\affiliation{Department of Physics and State Key
Laboratory of Nuclear Physics and Technology, Peking
University, Beijing, 100871, China}

\author{Chong Sheng Li\footnote{Electronic
address: csli@pku.edu.cn}}
\affiliation{Department of Physics and State Key
Laboratory of Nuclear Physics and Technology, Peking
University, Beijing, 100871, China}
\affiliation{Center for High Energy Physics, Peking
University, Beijing, 100871, China}

\author{Hua Xing Zhu \footnote{Present address: SLAC  National Accelerator Laboratory, Stanford University, Stanford, CA 94309, USA}}

\affiliation{Department of Physics and State Key
Laboratory of Nuclear Physics and Technology, Peking
University, Beijing, 100871, China}





\begin{abstract}
 \vspace*{0.3cm}
We study the factorization and resummation of t-channel top quark  transverse momentum distribution at large $p_T$
in the SM at both the Tevatron and the LHC with soft-collinear effective theory.
The cross section in the threshold region can be factorized into a convolution of hard, jet and soft functions.
In particular, we first calculate the NLO soft functions for this process, and give a RG improved cross section by evolving the different functions to a common scale.
Our results show that the resummation effects increase the NLO results by about $9\%\sim 13\%$ and $4\%\sim 9\%$
when the top quark $p_T$ is larger than $50$ and 70~GeV at the Tevatron and  the 8 TeV LHC, respectively.
Also, we discuss the scale independence of the cross section analytically,
and show how to choose the proper scales at which the perturbative expansion can converge fast.

\end{abstract}
\maketitle
\newpage



\section{Introduction}
\label{sec:1}

The top quark is the heaviest particle so far discovered, with a mass close to the electroweak symmetry breaking scale,
and closely related to various extensions of the standard model (SM).
Thus, it provides an effective probe for the electroweak symmetry breaking mechanism and a test for the predictions of the SM through its production or decay.

The production of the single top provides a good opportunity to study the charged weak current interactions of the top quark,
e.g., the structure of the $Wtb$ vertex~\cite{Bernreuther:2008ju}.
Besides, it is an important background in many new physics searches at hadron colliders.
However, due to the difficulties in discriminating its signature from the large background,
it has taken a long time after the discovery of the top quark for the D0~\cite{Abazov:2009ii} and CDF~\cite{Aaltonen:2009jj}
collaborations at the Tevatron to observe the single top production.
Recently, the ATLAS and CMS collaborations at the LHC have also measured the
cross section of the single top production at low integrated luminosities \cite{Aad:2012ux,:2012ep}.

Among the three production modes at hadron colliders,
the t-channel is especially important because of its largest cross section
at both the Tevatron and the LHC.
This process has been extensively studied, including the next-to-leading order (NLO) QCD corrections
based on the 2 $\to$ 2 leading order (LO) process, called the five-flavor (5F) scheme \cite{Bordes:1994ki,Stelzer:1995mi,Harris:2002md,Sullivan:2004ie,
Campbell:2004ch,Cao:2004ky,Cao:2005pq,Campbell:2009gj}.
It has been shown that the NLO corrections increase the LO cross section by about $9\%$  and $5\%$ at the Tevatron and  LHC,
respectively.
In Ref. \cite{Campbell:2009ss}, the NLO calculation of the t-channel production
based on the 2 $\to$ 3 LO process, called the four-flavor (4F) scheme, was presented, which shows
 that the inclusive cross section in the 4F scheme is smaller than in the 5F scheme while
the uncertainty in the 4F scheme is larger than in 5F scheme.
This is due to the fact that
in the 5F scheme, the large logarithm of the form log($Q^2/m_b^2$),
due to the initial bottom quark,  is resummed into the bottom quark
parton distribution functions (PDFs) and thus the scale dependence is significantly reduced.
Besides, in the 5F scheme, the parton shower Monte Carlo simulation for the t-channel single top production
was studied  \cite{Frixione:2005vw,Frixione:2008yi,Alioli:2009je},
and the threshold resummation for this process is
carried out with the conventional resummation method
~\cite{Kidonakis:2006bu,Kidonakis:2007ej,Kidonakis:2011wy},
where the partial next-to-next-to-next-to-leading order results are obtained by expanding the resummed cross sections
to avoid the infrared singularities and ambiguities from prescription dependence.

In this work, we investigate the resummation of the t-channel single top production in the 5F scheme
using soft-collinear effective theory (SCET)
~\cite{Bauer:2000ew,Bauer:2000yr,Bauer:2001ct,Bauer:2001yt,Becher:2006nr}.
SCET is developed to describe the behavior of the QCD interactions in collinear and soft regions
with the short distance information encoded in the Wilson coefficients.
It is very suitable to deal with the scattering processes with multiple scales.
In the past ten years, SCET has proved very useful in high energy hard scattering processes.
In general, these processes can be divided into two kinds, i.e., the timelike and spacelike.
The timelike processes produce a timelike particle in the intermediate or final state, including
Drell-Yan production~\cite{Idilbi:2005ky,Idilbi:2005er,Becher:2007ty,Stewart:2009yx},
Higgs boson production~\cite{Gao:2005iu,Idilbi:2005er,Ahrens:2008qu,Ahrens:2008nc,Zhu:2009sg,Mantry:2009qz},
$e^+e^-$ annihilation  to hadrons~\cite{Lee:2006nr,Fleming:2007qr,Fleming:2007xt,Bauer:2008dt,Schwartz:2007ib},
color-octet scalar production~\cite{Idilbi:2009cc},
direct top quark production via FCNC coupling \cite{Yang:2006gs},
and s-channel single top production \cite{Zhu:2010mr}.
The spacelike processes involve a spacelike particle in the intermediate state, such as
deep-inelastic scattering~\cite{Manohar:2003vb,Chay:2005rz,Becher:2006nr,Chen:2006vd},
direct photon production~\cite{Becher:2009th} and $W$ ($Z$) boson production at large transverse momentum $p_T$ \cite{Becher:2011fc}.
Note that some processes are a mix of these two kinds,
e.g., the top quark pair production~\cite{Ahrens:2009uz,Ahrens:2010zv,Beneke:2011mq}.

The threshold region can be easily defined for the timelike processes.
It is usually defined as the limit $z=m^2/s \to 1$, where $m$ is the invariant mass of the time-like particle
and $s$ is the square of the center-of-mass energy.
For the spacelike processes, the threshold region is a little more subtle.
The threshold region for the deep-inelastic scattering process is given by the Bjorken scaling variable $x \to 1$.
For the direct photon production and $W$ ($Z$) boson production at large $p_T$, the threshold region is
approached when $S_4=M^2 \to 0$, where $M$ is the mass of everything in the final state except the photon ($W$ or $Z$).
The t-channel single top production is a spacelike process involving four colored external particles.
We  define the threshold region as $S_4=P_X^2 \to 0$, similar to the case of $W$ or $Z$ production at large $p_T$,
where $P_X^2$ represents the mass squared of everything in the final state except the top quark.
In this threshold region, the cross section can be factorized as
\begin{equation}
    \sigma=H\otimes J \otimes \mathcal{S} \otimes f_{P_a} \otimes f_{P_b},
\end{equation}
where $H, J, \mathcal{S}, f_{P} $ are the hard function, jet function, soft function and PDF, respectively.
The hard function incorporates the short distance contributions arising from virtual corrections.
The jet function describes all collinear interactions inside the outgoing jet.
The soft gluon effects coming from all colored particles are contained in the soft function.
The PDF denotes the probability of finding a particular parton in the proton.

The final states of the t-channel single top production at hadron colliders
consist of a single top quark and a jet at the LO.
Additional soft gluons can be emitted from the colored initial and final state particles,
and collinear gluons can be emitted in the jet.
These contributions are of higher orders in $\alpha_s$, but can be large in the threshold region.
Besides, the hard part of this process receives a large correction since the usually chosen
renormalization scale and the typical transferred momentum are $m_t$ and $Q=\sqrt{-\hat{t}}$, respectively,
which yield a large logarithm of the form ${\rm ln}(m_t^2/Q^2)$.
Therefore, it is necessary to resum all of these  large logarithm to all orders.
In the SCET approach, the different scales in a process can be separated
because the soft and collinear degrees are decoupled by the redefinition of the fields \cite{Bauer:2001yt}.
At each scale, one only needs to deal with the relevant degrees of freedom.
In this way, reliable perturbative expansions can be achieved easily, and
the dependencies of the final results on the scales are well controlled by the renormalization group (RG) equations.
As a result, the singular terms in the hard, jet and soft functions can be resummed conveniently.
Furthermore,
in numerical calculations, we find that for top quark $p_T>$ 50 GeV, the singular terms approximate the fixed-order calculations well,
but for $p_T<$ 50 GeV, the singular terms do not dominate over the NLO corrections.
This is understood because in the larger $p_T$ region, the phase space for the additional emitted gluon is more constrained
so that the main contribution comes from the soft gluon effects.
Thus, we need to know an improved resummation prediction on the top quark transverse momentum distribution in the region of large $p_T$, instead of the total cross section.
Such a t-channel top quark transverse momentum distribution is actually an observable which can be compared directly with the experimental results
\footnote{We have discussed this with the ATLAS and D0 experimentalists by email. They plan to provide such differential distributions
in an update of the cross section measurement.}, and is an important background in searching for  new physics.
For example, if there is an extra gauge boson $W'$ with a mass around 1 TeV and the standard-model-like couplings,
it is better to search this gauge boson in the $t\bar{b}$ final states than the two light jets final states because of the large dijet background from QCD processes.
Moreover, one should concentrate on the events with large top quark $p_T$ since the top quark from the decay of $W'$ usually has a large momentum.
In this case, a precise knowledge of the t-channel top quark transverse momentum distribution at large $p_T$ in the SM is necessary.

This paper is organized as follows.
In Sec.~\ref{sec:scet}, we give a brief introduction to SCET.
In Sec.~\ref{sec:kine}, we analyze the kinematics of the t-channel single top production process and give the definition of the threshold region.
In Sec.~\ref{sec:form}, we present the factorization and resummation formalism
for the t-channel single top production in momentum space.
In Sec.~\ref{sec:nlo}, we calculate the hard and soft functions at NLO.
Then, we study the scale independence of the final result analytically.
In Sec.~\ref{sec:nume}, we discuss the numerical results for t-channel top quark transverse momentum distribution at the Tevatron and the LHC.
We conclude in Sec.~\ref{sec:conc}.

\section{Brief introduction to SCET}
\label{sec:scet} To describe collinear fields in SCET, it is convenient to define a lightlike vector
$n_\mu=(1,\mathbf{n}),\mathbf{n}^2=1$.
Any four-vector can be light-cone decomposed with respect to $n_\mu$ and $\bar{n}_\mu=(1,-\mathbf{n})$ as
\begin{equation}
 l^\mu = l^- \frac{n^\mu}{2} + l^+ \frac{\bar{n}^\mu}{2} +
l^\mu_{n\perp},
\end{equation}
with $l^+=n\mcdot l$ and $l^-=\bar{n}\mcdot l$.
The momentum of a collinear particle moving along the $n^\mu$ direction has the following scaling
\begin{equation}
 p^\mu = (p^+, p^-, p_{n\perp} )\sim (\lambda^2, 1,
\lambda),
\end{equation}
while for a soft particle, the momentum scales as
\begin{equation}
 q\sim (\lambda^2, \lambda^2, \lambda^2),
\end{equation}
where $\lambda\ll 1$ is a small expansion parameter in SCET.
For example, for an energetic jet with invariant mass $m_J$ and energy $E_J$, $\lambda=m_J/E_J$.
From the momentum scaling, one can see that the interaction between collinear fields of different directions
$n_i$ and $n_j$ with $n_i\cdot n_j \gg \lambda^2$ will inevitably change the momentum scaling;
thus it is forbidden in SCET, but can be included as an external current in our computation.
The soft fields, on the other hand, can interact with any collinear field without changing the scaling.

In SCET, the $n$-collinear quark $\psi_n$ and gluon field $A_n^{\mu}$ can be written as
\begin{eqnarray}
\label{colfield}
 \chi_n(x) &=& W^\dagger_n (x) \xi_n(x),\quad \xi_n(x)=\frac{\nslash\bnslash}{4}\psi_n(x),
\nn
\\
\mathcal{A}^\mu_{n\perp}(x) &=& \left[ W^\dagger_n
iD^\mu_{n\perp}W_n(x) \right],
\end{eqnarray}
where
\begin{equation}
 iD^\mu_{n\perp} = \mathcal{P}^\mu_{n\perp}+g_s
A^\mu_{n\perp}
\end{equation}
is the collinear covariant derivative and the label operator $\cal P$ is defined to project out the large momentum component of the collinear field,
e.g., ${\cal P}^\mu_n \xi_n = \bar{p}^\mu \xi_n$.
Here we have split $p$ into a sum of large label momentum and small residue momentum,
\begin{equation}
 p^\mu = \bar{p}^\mu + k^\mu, \qquad \text{with}\qquad
\bar{p}^\mu = p^-\frac{n^\mu}{2} + p^\mu_{n\perp}.
\end{equation}
The $n$-collinear Wilson line,
\begin{equation}
 W_n(x) = \mathbf{P} \exp \left( ig_s \int^0_{-\infty} ds\,
\bn \mcdot A^a_n(x+s\bn)t^a \right),
\end{equation}
which describes the emission of arbitrary $n$-collinear gluons from an $n$-collinear quark or gluon,
is constructed to make the collinear fields as defined in Eq.~(\ref{colfield}) invariant under the collinear gauge transformation.
The operator $\mathbf{P}$ is the path-ordered operator acting on the color generator $t^a$.

At the LO in $\lambda$, only the $n\mcdot A_s$ component of soft gluons can interact with the $n$-collinear field.
Such interaction is eikonal and can be removed by a field redefinition~\cite{Bauer:2001yt}:
\begin{eqnarray}
\label{eqs:frd}
 \chi_n(x) & \to & Y_n(x_{-})\chi_n(x),
\nn
\\
 \mathcal{A}^\mu_{n\perp}(x) &\to & Y_n(x_{-})
\mathcal{A}^{\mu}_{n\perp}(x) Y^\dagger_n(x_{-}),
\end{eqnarray}
where
\begin{equation}
 Y_n(x) = \mathbf{P} \exp\left( ig_s\int^0_{-\infty}ds\,
n\mcdot A^a_s(x+sn)t^a\right)
\end{equation}
for an incoming Wilson line~\cite{Bauer:2001yt,Chay:2004zn}.
And for an outgoing Wilson line, it is defined as
\begin{equation}
 \tilde{Y}_n(x) = \mathbf{P} \exp\left(
-ig_s\int^\infty_{0}ds\, n\mcdot A^a_s(x+sn)t^a\right).
\end{equation}
The soft gluon fields are multipole-expanded around $x_{-}$ to maintain a consistent power counting in $\lambda$.
For the interaction between the soft gluon fields and massive quark fields, there exists a similar timelike Wilson line \cite{Korchemsky:1991zp}, for example,
\begin{equation}
 Y_v(x) = \mathbf{P} \exp\left( ig_s\int^0_{-\infty}ds\,v\mcdot A^a_s(x+sv)t^a\right).
\end{equation}
After the field redefinition, the LO SCET Lagrangian is factorized into a sum of different collinear sectors and a soft sector,
which do not interact with each other:
\begin{equation}
\label{fl}
 \mathcal{L}_{\rm SCET} = \sum_{n_i} \mathcal{L}^{(0)}_{n_i}
+ \mathcal{L}_s + \cdots.
\end{equation}
The decoupling of soft gluons from collinear fields is crucial for deriving the factorization formula.

\section{Analysis of kinematics}
\label{sec:kine}

In this section, we introduce the relevant kinematical variables needed in our analysis.
As an example, we consider the subprocess
\begin{equation}
    u (p_a)+ b (p_a) \to t (q) + X,
\end{equation}
\begin{figure}
  \includegraphics[width=0.3\linewidth]{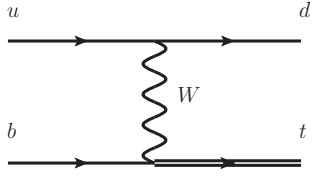}\\
  \caption{The Feynman diagram for the subprocess $ub\to dt$.}
  \label{fig:lo}
\end{figure}
whose Feynman diagram is shown in Fig. \ref{fig:lo}.
First of all, we define two lightlike vectors along the beam directions, $n_a$ and $n_b$, which are related by $n_a=\bar{n}_b$.
Then we introduce initial collinear fields along $n_a$ and $n_b$ to describe the collinear particles in the beam directions.
In the center-of-mass frame of the hadronic collision, the momenta of the incoming hadrons can be written as
\begin{equation}
 P^\mu_a=E_{\rm CM}\frac{n^\mu_a}{2},\qquad P^\mu_b=E_{\rm
CM}\frac{n^\mu_b}{2}.
\end{equation}
Here $E_{\rm CM}$ is the center-of-mass energy of the collider and we have neglected the masses of the hadrons.
The momenta of the incoming partons, with a light-cone momentum fraction of the hadronic momentum, are
\begin{equation}
 p_a = x_a E_{\rm CM}\frac{n^\mu_a}{2},\qquad
p_b= x_b E_{\rm
CM}\frac{n^\mu_b}{2}.
\end{equation}
At the hadronic and partonic levels, the momentum conservation means
\begin{equation}
 P_a + P_b = q + P_X,
\end{equation}
and
\begin{equation}
 p_a+p_b=q + p_X,
\end{equation}
respectively, where $q$ is the momentum of the top quark.
We define the partonic jet with  momentum $p_X$, which represents the set of all final state partons except the top quark in the partonic processes,
while the hadronic jet with  momentum $P_X$ contains all the hadrons as well as the beam remnants in the final state, except the top quark.
Explicitly, $p_X=p_1 + k$, where $p_1$ is the momentum of the final state collinear partons forming the jet and $k$ is the momentum of the soft radiations.
Such division of momentum is artificial and we have to integrate over the soft momentum to obtain a physical observable.

We also define the Mandelstam variables as
\begin{equation}
 s=(P_a+P_b)^2,\quad u=(P_a-q)^2, \quad t=(P_b-q)^2
\end{equation}
for hadrons, and
\begin{equation}
  \hat{s}=(p_a+p_b)^2,\quad
\hat{u}=(p_a-q)^2, \quad \hat{t}=(p_b-q)^2
\end{equation}
for partons, respectively. In terms of the Mandelstam variables, the hadronic and partonic threshold variables are defined as
\begin{eqnarray}
S_4 &\equiv& P^2_X = s+t+u-m^2_t,  \\
s_4 &\equiv& p^2_X =\hat{s}+\hat{t}+\hat{u}- m^2_t,
\end{eqnarray}
where $m_t$ is the mass of top quark. The hadronic threshold limit is defined as $S_4 \to 0$~\cite{Laenen:1998qw}.
In this limit, the final state radiations and beam remnants are highly suppressed,
which leads to final states consisting of a top quark and a narrow jet, as well as the remaining soft radiations.
Taking this limit requires $x_a\to 1,\;x_b\to 1,\; s_4 \to 0$ simultaneously.
In this limit, we get
\begin{eqnarray}\label{eqs:s4}
  S_4 &=& s_4 + \hat{s}(\frac{1}{x_ax_b}-1)+(\hat{t}-m^2_t)(\frac{1}{x_b}-1)+(\hat{u}-m^2_t)(\frac{1}{x_a}-1) \nn\\
   &\approx& s_4+\hat{s}(\bar{x}_a+\bar{x}_b)+(\hat{t}-m^2_t)\bar{x}_b+(\hat{u}-m^2_t)\bar{x}_a \nn\\
   &\approx& s_4+(-\hat{t})\bar{x}_a+(-\hat{u})\bar{x}_b,
\end{eqnarray}
where $\bar{x}_{a,b}=1-x_{a,b}$.
This expression can help to check the factorization scale invariance which is shown with more detail in the following.
The hadronic threshold enforces the partonic threshold.
However, the reverse is not true. The partonic threshold $s_4 \to 0$ does not forbid a significant amount of beam remnants.
We note that in both hadronic and partonic threshold limits, the top quark is not forced to be produced at rest;
$i.e.$ it can have a large momentum.
For later convenience, we can also write the threshold variable as
\begin{equation}
\label{eqs:s4a}
 s_4=p^2_X=(p_a+p_b-q)^2=p^2_1+2k^+
E_1+\cO(k^2),
\end{equation}
where $k^+=n_1\mcdot k$, $k$ is the sum of the momenta of soft radiations;
$E_1$ is the energy of the quark jet and $n_1$ is the lightlike vector associated with the jet direction.
In the threshold limit~($s_4\to 0$), incomplete cancellation between real and virtual corrections leads to singular distributions
$\alpha^n_s
[\ln^m(s_4/m^2_t)/s_4]_+$, with $m \leq 2n-1$.
It is the purpose of threshold resummation to sum up these contributions to all orders in perturbation theory.

The inclusive total cross section of the t-channel single top production can be written as
\begin{eqnarray}\label{eqs:main}
  \sigma &=& \int dx_a \int dx_b \int d\hat{t} \int d\hat{u} f_{i/P_a}(\mu_F,x_a)f_{j/P_b}(\mu_F,x_b)
  \frac{1}{2\hat{s}}\frac{d\hat{\sigma}_{ij}}{d\hat{t}d\hat{u}} \nn\\
   &=&\int_{0}^{p^2_{T,max}} dp_T^2 \int_{-y_{max}}^{y_{max}} dy \int_{x_{b,min}}^1 dx_b \int_0^{s_4^{max}} ds_4
   \frac{1}{2(x_bs+u-m_t^2)} f_{i/P_a}(\mu_F,x_a)f_{j/P_b}(\mu_F,x_b)
   \frac{d\hat{\sigma}_{ij}}{d\hat{t}d\hat{u}},\nn\\
\end{eqnarray}
where we have changed the integration variables to be the top quark transverse momentum squared $p^2_T$, rapidity $y$, $x_b$ and $s_4$.
The regions of the integration variables are limited by
\begin{eqnarray}
  p^2_{T,max} &=& \frac{(s-m_t^2)^2}{4s}, \nn\\
  y_{max} &=& \frac{1}{2}{\rm ln} \frac{1+\sqrt{1-sq}}{1-\sqrt{1-sq}}, {\rm ~with~ }sq=\frac{4s(p_T^2+m_t^2)}{(s+m_t^2)^2}, \nn \\
  x_{b,min} &=& \frac{-u}{s+t-m_t^2}, \nn\\
  s_4^{max} &=& x_b(s+t-m_t^2)+u,
\end{eqnarray}
with
\begin{eqnarray}
  t &=& m_t^2 -\sqrt{s}\sqrt{p_T^2+m_t^2}e^{y} \nn \\
  u &=& m_t^2 -\sqrt{s}\sqrt{p_T^2+m_t^2}e^{-y}.
\end{eqnarray}
The other kinematical variables can be expressed in terms of these four integration variables.

\section{Factorization and Resummation Formalism}
\label{sec:form}

To derive a factorization formula for the t-channel single top production in SCET,
we first have to match the full QCD onto the effective theory.
In this section, we follow the convention and formalism in~\cite{Bauer:2008jx,Bauer:2010vu},
where the matching is performed in momentum space.
The relevant operator in QCD responsible for the t-channel single top production is
\begin{equation}
    O(x)=(\bar{d}\gamma^{\mu}P_L u\bar{t}\gamma_{\mu}P_L b)(x),
\end{equation}
where we have adopted the Feynman gauge for the $W$ boson propagator.
This operator contains three massless quarks,
which can be described by collinear quarks in SCET, and a massive quark,
which can be described by heavy quark effective theory \cite{Isgur:1989vq}.
The presence of three different directions and a massive quark is a feature of single top production at hadron colliders.
The operator $O(p)$, which is the Fourier transform of $O(x)$, can be written in terms of the momentum-space SCET fields
in the threshold region as
\begin{eqnarray}
\label{eqs:ope}
 O(p) &=& \int\,
\frac{d^4p_a}{(2\pi)^4} \frac{d^4p_b}{(2\pi)^4} \frac{d^4p_1}{(2\pi)^4}
\frac{d^4p_2}{(2\pi)^4} \frac{d^4k_s}{(2\pi)^4} \mathcal{C}_I(p_a,p_b;p_1,p_2)
\nn\\
&&\times \cO_{in}(p_a,p_b) \cO_{out} (p_1,p_2)
\cO_{S,I}(k_s)(2\pi)^4\delta^{(4)}(p-p_a-p_b+p_1+p_2+k_s),
\end{eqnarray}
where the operator $\cO_{in}(p_a,p_b)$ is responsible for annihilating the initial u and b quarks with momenta $p_a$ and $p_b$, respectively, or explicitly,
\begin{equation}
\label{isf}
\cO^{cd}_{\alpha\beta,in} = \chi^c_\alpha(p_a)\chi^d_\beta(p_b),
\end{equation}
and $\cO_{out} (p_1,p_2)$ is responsible for creating the final d and t quarks with momenta $p_1$ and $p_2$, respectively, or explicitly,
\begin{equation}
\label{fsf}
 \cO^{ef}_{\gamma\delta,out} = \bar{\chi}^f_\delta (p_1)\bar{h}^e_{\gamma,v}(p_2) .
\end{equation}
Note that we have described the top quark in terms of the heavy quark effective field with a label velocity $v$~\cite{Isgur:1989vq}.
Since there are two fermion lines in this process,
either of which connects an initial and a final states,
we indicate them  with the Lorentz ($\alpha,\beta,\gamma,\delta$) and color indices ($c,d,e,f$) explicitly,
and retain only quark fields in the operators for simplicity,
leaving the other structure in the matching coefficient $\mathcal{C}_I$,
which is at the LO level
\begin{equation}
\label{lohmc}
\mathcal{C}_I^{\delta\alpha,\gamma\beta}=
i\frac{g^2 V_{ud}V_{tb}}{8(\hat{t}-M^2_W)}
(\gamma^\mu(1-\gamma^5))^{\delta\alpha}
(\gamma_\mu(1-\gamma^5))^{\gamma\beta} \delta_{I1}.
\end{equation}
Here, the electroweak coupling is defined by $g^2=8G_F M_W^2/\sqrt{2}$ where $G_F$ is the Fermi constant.
$ V_{ij}$ is the CKM matrix element and $M_W$ is the mass of the $W$ boson.
$\delta_{I1}$ denotes the color structure of the t-channel
single top production at LO in the singlet-octet basis
\begin{equation}
 |c_1\ra = \delta_{fc}\delta_{ed},\qquad
 |c_2\ra=(t^a)_{fc}(t^a)_{ed},
\end{equation}
where $t^a$ is the generator of the gauge group $SU(3)_c$, satisfying
\begin{equation}
{\rm tr}[t^at^b]=\frac{1}{2}\delta^{ab}.
\end{equation}
$I=1$ or $2$ is an index in this color space.
In Eq. (\ref{eqs:ope}), we can separate the soft gluon field
from collinear or massive fields because of the field redefinition in Eq. (\ref{eqs:frd}).

The soft operators $\cO_{S,I}$, which are responsible for the soft
interactions between different collinear sectors and the top quark, are
expressed as
\begin{eqnarray}
 \cO_{S,1}^{fced}(k_s)&=&\int\,d^4x e^{-ik_s\,\mcdot \,x}
 \mathbf{T}\left[ \left(\tilde{Y}^\dagger_{n_1} (x)  Y_{n_a}(x) \right)^{fc}
 \left(\tilde{Y}^\dagger_{v}(x)Y_{n_b} (x)\right)^{ed}\right],\nn\\
 \cO_{S,2}^{fced}(k_s)&=&\int\,d^4x e^{-ik_s\,\mcdot \,x}
 \mathbf{T}\left[ \left(\tilde{Y}^\dagger_{n_1} (x) t^a Y_{n_a}(x) \right)^{fc}
 \left(\tilde{Y}^\dagger_{v}(x) t^a Y_{n_b} (x)\right)^{ed}\right],
\end{eqnarray}
where the time-ordering operator $\mathbf{T}$ is required to ensure
the proper ordering of soft gluon fields in the soft Wilson line.

The total cross section for t-channel single top production in the threshold region can be written as
\begin{eqnarray}
\label{eq1}
\sigma &=& \frac{1}{2s}
\sum_X \la I | O^\dagger_x(0) | X \ra \la X|
O_x(0)
| I\ra (2\pi)^4\delta^4(P_a+P_b-q-P_X)
\nn
\\
&=& \frac{1}{2s}
\sum_X \int d^4x\la I | O^\dagger_x(x) | X \ra
\la X|
O_x(0)
| I\ra
\nn
\\
&=& \frac{1}{2s}
\sum_X
\int d^4x\int \frac{d^4 k}{(2\pi)^4} e^{-ik\cdot x} \int
\frac{d^4 p}{(2\pi)^4}
\la I | O^\dagger(k) | X \ra \la X|
O(p)
| I\ra
\nn
\\
&=& \frac{1}{2s} \sum_X
\int\,\frac{d^4p}{(2\pi)^4}
\la I| O^\dagger(0) | X \ra \la X| O(p)| I
\ra ,
\end{eqnarray}
where $|I\ra=|P_a P_b\ra$ denotes the initial state protons
(antiprotons). Here we distinguish the position space operator from
the momentum space one by a subscript $x$. The restriction on the
sum over final states $|X\ra$ is that the final state
configuration consists only of a top quark jet whose 3-momentum is
in the direction of $\bar{n}_1$, a d-quark jet in the
direction of $n_1$, and soft radiations. This is the configuration
that is relevant to threshold resummation and that we are interested in.
Under this condition the final state can be written as $|X\ra = |X_t
X_1 X_s\ra$, where $|X_t\ra$, $|X_1\ra$ and $|X_s\ra$ denote the top
quark jet, the d-quark jet and the remaining soft radiations,
respectively. In the second line of Eq.~(\ref{eq1}), we have used
the momentum conservation delta function to shift the operator
$O_x^\dagger$ to point $x$, and in the third line we have written
the operators in momentum
space, which are matched onto SCET operators.

Using the notation $\Phi_2=\{p_a,p_b;p_1,p_2\}$ to express a phase
space point~\cite{Bauer:2008jx} with
$d\Phi_2=d^4 p_a d^4 p_b d^4p_1 d^4p_2/(2\pi)^{16}$
and $\Phi_2 - k_s = p_a + p_b - p_1 - p_2 - k_s$,  we can write
Eq.~(\ref{eq1}) in a compact form
\begin{eqnarray}
\label{fdcs}
  \sigma&=& \frac{1}{2s} \sum_X
\int d\Phi'_2 d\Phi_2 \mathcal{C}^*_J(\Phi'_2) \mathcal{C}_I(\Phi_2)
\int \frac{d^4k'_s}{(2\pi)^4}\,\frac{d^4k_s}{(2\pi)^4}
(2\pi)^4\delta^{(4)}(\Phi_2-k_s) \nn \\
&& \times
\la I | (\cO'_{in}\cO'_{out}\cO'_{S,J})^\dagger |X_t X_1 X_s\ra
\la X_t X_1 X_s| (\cO_{in}\cO_{out}\cO_{S,I}) |I\ra.
\end{eqnarray}
As we mentioned before, different collinear sectors are
decoupled due to field redefinition,
and thus the matrix element in Eq.~(\ref{fdcs}) can be
factorized into a product of several matrix elements,
which obey certain RG equations.

In the following, we further show the matrix elements mentioned above.
First, we deal with the top quark sector.
Since we have decoupled the soft interaction by field redefinition,
the top quark now should be regarded as a noninteracting particle,
which can be written as
\begin{eqnarray}
 &&\sum_{X_t}\int \frac{d^4 p'_2}{(2\pi)^4}\frac{d^4 p_2}{(2\pi)^4}
 \la 0| {h^{e'}_{\gamma',v'}} (p'_2) |X_t\ra \la X_t| \bar{h}^e_{\gamma,v}(p_2) | 0 \ra \nn \\
&=& \int\frac{d^3 q}{2 E_q(2\pi)^3}(\Slash{q} + m_t)_{\gamma'\gamma}\delta^{e'e},
\end{eqnarray}
where summation over the final state $|X_t\ra$ gives rise to a top quark phase space integral.
Next, we define the soft function by the soft matrix element as
\begin{eqnarray}\label{eqs:sft}
  \int dk^+ S^{d'e'c'f'fced}_{JI}(k^+,\mu) &=& \frac{1}{N_c^2}\sum_{X_s}
 \int dk^+ \frac{d^4 k'_s}{(2\pi)^4}\frac{d^4
k_s}{(2\pi)^4} \la 0|{\mathcal{O}^{\dagger,d'e'c'f'}_{S,J}}(k'_s) |X_s\ra\nn\\
&&\la X_s |\mathcal{O}^{fced}_{S,I} (k_s)|0 \ra\delta(k^+ -n_1\mcdot k_s),
\end{eqnarray}
where $N_c$ is the number of colors and we have inserted into the above equation an identity
operator
\begin{equation}
 \mathbf{1}=\int dk^+ \, \delta[k^+ - n_1 \mcdot k_s],
\end{equation}
because of the constraint from Eq.~(\ref{eqs:s4a}), which
expresses the multipole expansion of a soft field
interacting with a collinear field~\cite{Becher:2009th}.
Note that the summation over a final state can be performed
$\sum_{X_s} |X_s(k'_s)\ra \la X_s(k_s)|=(2\pi)^4\delta^{(4)}(k'_s-k_s)$
since there is no
restriction in the summation and also there is no explicit dependence of the final states on
$|X_s\ra$.
Since we are only interested in  the cross sections at large top quark $p_{T}$,
the final state top quark, jet function and
PDFs can be considered to be diagonal in color space. Then
we can contract their color indices to obtain
the soft function matrix
\begin{equation}\label{eqs:softmatrix}
  S_{JI}(k^+,\mu) = \delta^{f'f}\delta^{c'c} \delta^{e'e} \delta^{d'd} S^{d'e'c'f'fced}_{JI}(k^+,\mu).
\end{equation}
At the LO, it can be written as
\begin{equation}\label{eqs:softLO}
 \mathbf{S}(k^+,\mu)=\delta(k^+)\frac{1}{N_c^2}
\left(
\begin{array}{cc}
C_A^2 & 0\\
0 & \frac{C_A^2-1}{4}
\end{array}
\right),
\end{equation}
where $C_A$ is the Casimir operator for the adjoint representation of $SU(3)_c$.
At the NLO, the calculation of the soft function boils down to
the evaluation of eikonal diagrams~\cite{Becher:2009th}.
Since the virtual corrections in SCET vanish,
only real emission diagrams are needed to be evaluated.
The details of the calculation of these diagrams are given in Appendix \ref{sec:soft}.

For the final state d-quark jet sector, we have
\begin{eqnarray}
\label{eqs:jet}
&& \sum_{X_1}\int \frac{d^4p'_1}{(2\pi)^4}\frac{d^4p_1}{(2\pi)^4}
\la 0| \chi^{f'}_{\delta'}(p'_1) |X_1\ra \la X_1| \bar{\chi}^f_\delta (p_1) |0\ra \nn \\
&&=\delta^{f'f}\int \frac{d^4p_1}{(2\pi)^3}\left(\frac{\Slash{n}_1}{2}\right)_{\delta\delta'}\theta(p^0_1) \bar{n}_1 \cdot p_1 J(p^2_1),
\end{eqnarray}
where the summation over the collinear state has been performed and $J$ is the spin- and color-singlet jet function,
defined as
\begin{equation}
\theta(p^0)\bar{n}_1 \cdot p J(p^2)=\frac{1}{8\pi N_c}
\int \frac{d^4p'}{(2\pi)^4} {\rm Tr}
\la 0| \bnslash_1  \chi(p')  \bar{\chi} (p) |0\ra,
\end{equation}
where Tr represents the trace over spin and color indices. At LO, it is just $\delta(p^2)$.
Finally, the initial state $n_a$ collinear sector reduces to the
conventional PDFs:
\begin{eqnarray}
 &&\int  \frac{d^4p'_a}{(2\pi)^4}
\frac{d^4p_a}{(2\pi)^4} \la P_a | \bar{\chi}^{c'}_{\alpha'} (p'_a)  \chi^c_\alpha (p_a) | P_a \ra
= \frac{1}{2 N_c}\delta^{c'c} \int^1_0 \frac{dx_a}{x_a}
\left(x_a E_{\rm CM}\frac{\Slash{n}_a}{2}\right)_{\alpha \alpha'} f_{u/P_a}( x_a, \mu)
\end{eqnarray}
and similarly for the matrix element for $n_b$ direction.
Thus the momenta of incoming partons are given by
$p_{a(b)}=x_{a(b)}E_{\rm CM}n_{a(b)}/2$.

Combining the above expressions, we obtain (up to power corrections)
\begin{eqnarray}
\sigma &=&\int dx_a dx_b d\hat{t}d\hat{u} \frac{1}{2\hat{s}}
f_{i/P_a}(x_a,\mu) f_{j/P_b}(x_b,\mu)
\frac{d\hat{\sigma}_{ij}^{\rm thres}}{d\hat{t}d\hat{u}},\label{eqs:facmain0}
\end{eqnarray}
with
\begin{eqnarray}
\frac{d\hat{\sigma}_{ij}^{\rm thres}}{d\hat{t}d\hat{u}} &=&
\frac{1}{4N^2_c} \frac{1}{8\pi}\frac{1}{\hat{s}}
 \lambda_{0,ij} H_{IJ}(\mu) \nn \\
&& \times  \int dk^+\int dp_1^2 \, S_{JI}(k^+,\mu) J(p_1^2,\mu)\delta(s_4-p_1^2-2k^+E_1),
\label{eqs:facmain}
\end{eqnarray}
and
\begin{equation}
 \lambda_{0,ij} =  g^4|V_{id}|^2|V_{jt}|^2
 \frac{(\hat{s}-m^2_t)\hat{s}}{(\hat{t}-M^2_W)^2}.
\end{equation}
All the objects in the factorized Eq. (\ref{eqs:facmain0}) have precise field-theoretic definitions
so that they can be calculated directly  and systematically, except the nonperturbative PDF.
The convolution between the jet and soft functions suggests that the partonic threshold consists of two parts.
In the case of $s_4 = 0$, there are no collinear or soft gluons emitted.
In the small $s_4$ region, the number and momentum of collinear and soft gluons are constrained.

At the LO, the hard function $H_{IJ}$ is normalized to
$\delta_{I1}\delta_{J1}$.
In general, it is related to the amplitudes of the full theory, and is given by~\cite{Ahrens:2010zv}
\begin{eqnarray}
 \lambda_{0,ij} H^{(0)}_{IJ} &=& \frac{1}{\la c_I|c_I\ra \la
c_J
| c_J \ra} \la c_I | \mathcal{M}^{(0)}_{\rm ren} \ra \la
\mathcal{M}^{(0)}_{\rm ren} | c_J \ra,
\nn
\\
\lambda_{0,ij} H^{(1)}_{IJ} &=& \frac{1}{\la c_I|c_I\ra \la
c_J |
c_J \ra} \left(\la c_I | \mathcal{M}^{(1)}_{\rm ren} \ra
\la
\mathcal{M}^{(0)}_{\rm ren} | c_J \ra+\la c_I |
\mathcal{M}^{(0)}_{\rm ren} \ra \la
\mathcal{M}^{(1)}_{\rm ren} | c_J \ra\right),
\label{eqs:hardmatrix}
\end{eqnarray}
where $|\mathcal{M}_{\rm ren} \ra$ are obtained by
subtracting the IR divergences in the $\overline{\rm MS}$ scheme
from the UV  renormalized amplitudes of the full theory.

Because of the
special color structure of this process, the
hard function matrix elements do not contribute to the cross section
except for $H_{11}$ at the NLO level. In SCET, there is a RG
evolution factor connecting the hard scale $\mu_{h}$ and the final
common scale $\mu$, which would contain contributions from
nondiagonal elements  beyond NLO.
However, these nondiagonal contributions involve the gluon
connecting two fermion lines, resulting in a suppressed color factor $1/N_c^2$,
compared to diagonal ones.
Thus, we expect their contributions are small and can be neglected safely.
Then the t-channel single top production is considered to be
a double deep-inelastic-scattering (DDIS) process~\cite{Harris:2002md}.
In this case the hard function
$H_{11}$ can be further factorized into two parts, i.e.,~$H_{up}$ and
$H_{dn}$, which represent contributions from the up and down
fermion lines, respectively, in the Feynman diagram as shown in Fig. \ref{fig:lo}.
This separation is also helpful to make a reliable perturbative
prediction for the hard function.
The reason is that usually the loop corrections from the up and down
fermion lines contain large logarithms of the forms ln$(-\hat{t}/\mu_h^2)$
and ln$((-\hat{t}+m_t^2)/m_t/\mu_h)$, respectively; see Eqs. (\ref{hard_function})-(\ref{hardf_function}).
It is hard to choose a proper hard scale to make both of them small.
In the case of a DDIS process, the two separate hard parts
can be evaluated in different scales such that the
perturbative expansion is reliable in both parts.
As a consequence, we can rewrite Eq. (\ref{eqs:facmain}) as
\begin{eqnarray}\label{eqs:facmain2}
\frac{d\hat{\sigma}_{ij}^{\rm thres}}{d\hat{t}d\hat{u}} &=&
\frac{1}{4N^2_c} \frac{1}{8\pi}\frac{1}{\hat{s}}
 \lambda_{0,ij} H_{up}(\mu) H_{dn}(\mu) \nn \\
&& \times  \int dk^+\int dp_1^2 \, \mathcal{S}(k^+,\mu) J(p_1^2,\mu)\delta(s_4-p_1^2-2k^+E_1),
\end{eqnarray}
where $\mathcal{S}(k^+,\mu)$ denotes the component $S_{11}(k^+,\mu) $ in Eq. (\ref{eqs:softmatrix}).

\section{The Hard, Jet and Soft Functions at NLO }
\label{sec:nlo}
The hard, jet and soft functions describe interactions at different scales, and
they can be calculated order by order in perturbative theory at each scale.
At the next-to-next-to-leading logarithmic accuracy, we need the explicit expressions of hard, jet and soft functions up to NLO.

\subsection{Hard functions}
\label{subsec:41}
The hard functions are the absolute value squared of the Wilson coefficients of the operators,
which can be obtained by matching the full theory onto SCET.
In practice, we need to calculate the one-loop on-shell Feynman diagrams of this process in both the full theory and SCET.
In dimensional regularization, the facts that the IR structure of the full theory and the effective theory are identical
and that the on-shell integrals are scaleless and vanish in SCET imply that the IR divergence of the full theory is
just the negative of the UV divergence of SCET.
\begin{figure}
  \includegraphics[width=0.5\linewidth]{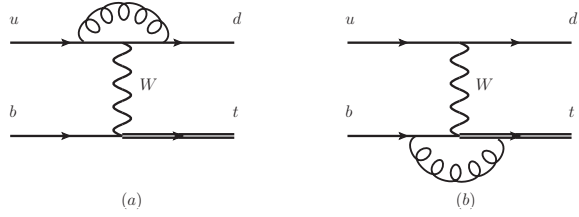}\\
  \caption{The one-loop Feynman diagrams for the t-channel single top production.}
  \label{fig:virtual}
\end{figure}
After calculating the one-loop on-shell Feynman diagrams, as shown in Fig. \ref{fig:virtual},
we get the hard functions at NLO as follows:
\begin{eqnarray}
\label{hard_function}
H_{up}(\mu_{h,up})&=&1+\frac{C_{F}\alpha_{s}(\mu_{h,up})}{4\pi}\biggl(
-2\mathrm{ln}^2\frac{-\hat{t}}{\mu_{h,up}^{2}}+6\mathrm{ln}\frac{-\hat{t}}{\mu_{h,up}^{2}}+c^{H,up}_1 \biggr),\\
\label{hardf_function}
H_{dn}(\mu_{h,dn})&=&1+\frac{C_{F}\alpha_{s}(\mu_{h,dn})}{4\pi}\biggl(
-4\mathrm{ln}^2\frac{-\hat{t}+m_t^2}{\mu_{h,dn}m_t}+10\mathrm{ln}\frac{-\hat{t}+m_t^2}{\mu_{h,dn}m_t}+c^{H,dn}_1\biggr),
\end{eqnarray}
where
\begin{eqnarray}
c^{H,up}_1&=&-16+\frac{\pi^2}{3},\\
c^{H,dn}_1&=&-\frac{2}{\lambda}~\mathrm{ln}(1-\lambda)+2\mathrm{ln}^2(1-\lambda)+6~\mathrm{ln}(1-\lambda)+4\mathrm{Li}_{2}(\lambda)-12-\frac{\pi^2}{6}\nn\\
&&+\frac{2m_t^2\hat{u}}{\hat{t}(\hat{s}-m_t^2)}\mathrm{ln}\frac{m_t^2}{m_t^2-\hat{t}},
\end{eqnarray}
with $\lambda=\hat{t}/(\hat{t}-m_t^2)$.
These results agree with those in Ref. \cite{Harris:2002md}.
In order to avoid large logarithms, the natural choices of $\mu_{h,up}$ and $\mu_{h,dn}$
are $\sqrt{-\hat{t}}$ and $(-\hat{t}+m_t^2)/m_t$, respectively.
The hard functions at the other scales can be obtained by evolution of RG equations.
The RG equations for hard functions are governed by the anomalous-dimension matrix, the structure of which
has been predicted up to four-loop and two-loop level for the case involving
massless \cite{Ahrens:2012qz} and massive partons  \cite{Becher:2009kw}, respectively.
In our case, we can write the RG equations for hard functions as
\begin{eqnarray}\label{eqs:hardupRG}
\frac{d}{d~\mathrm{ln}\mu_{h,up}}H_{up}(\mu_{h,up})&=&\biggl( 2\Gamma_{\rm cusp}~\mathrm{ln}\frac{-\hat{t}}{\mu_{h,up}^2}+2\gamma_{up}^V \biggr )H_{up}(\mu_{h,up}),\\
\label{eqs:harddnRG}
\frac{d}{d~\mathrm{ln}\mu_{h,dn}}H_{dn}(\mu_{h,dn})&=&
\biggl(2\Gamma_{\rm
cusp}~\mathrm{ln}\frac{-\hat{t}+m_t^2}{\mu_{h,dn}m_t}+2\gamma_{dn}^V
\biggr )H_{dn}(\mu_{h,dn}),
\end{eqnarray}
where $\Gamma_{\rm cusp}$ is related to the cusp anomalous dimension
of Wilson loops with lightlike segments~\cite{Korchemskaya1992169},
while $\gamma_{up}^V$ and $\gamma_{dn}^V$ control the single-logarithmic evolution.
Their expressions up to two-loop level are shown in Appendix \ref{sec:anodim}.

After solving the RG equations, we get the hard functions at an
arbitrary scale $\mu$:
\begin{eqnarray}
\label{eqs:hardup}
H_{up}(\mu)&=& \mathrm{exp}\bigl[ 4S(\mu_{h,up},\mu)-2a_{up}^V(\mu_{h,up},\mu)\bigr]\bigg( \frac{-\hat{t}}{\mu_{h,up}^2} \bigg)^{-2a_{\Gamma}(\mu_{h,up},\mu)}H_{up}(\mu_{h,up}),\\
\label{eqs:harddn}
H_{dn}(\mu)&=& \mathrm{exp}\bigl[ 2S(\mu_{h,dn},\mu)-2a_{dn}^V(\mu_{h,dn},\mu)\bigr]\bigg( \frac{-\hat{t}+m_t^2}{\mu_{h,dn}m_t} \bigg)^{-2a_{\Gamma}(\mu_{h,dn},\mu)}H_{dn}(\mu_{h,dn}),
\end{eqnarray}
where $S(\mu_{h,up},\mu)$ and $a_{up}^V$ are defined as~\cite{Becher:2006mr}
\begin{eqnarray}
S(\mu_{h,up},\mu)&=&-\int_{\alpha_s(\mu_{h,up})}^{\alpha_s(\mu)}d\alpha \frac{\Gamma_{\rm cusp}(\alpha)}{\beta(\alpha)}
 \int_{\alpha_s(\mu_{h,up})}^{\alpha}\frac{d\alpha^{\prime}}{\beta(\alpha^{\prime})},\\
a_{up}^V(\mu_{h,up},\mu)&=&-\int_{\alpha_s(\mu_{h,up})}^{\alpha_s(\mu)}d\alpha
\frac{\gamma_{up}^V(\alpha)}{\beta(\alpha)}.
\end{eqnarray}
$S(\mu_{h,dn},\mu)$, $a_{\Gamma}$ and $a_{dn}^V$ have similar expressions.

\subsection{Jet function}
The jet function $J(p^2,\mu)$, defined in Eq. (\ref{eqs:jet}), describes a quark jet of invariant mass squared $p^2$.
It is process independent and has been calculated at NLO in \cite{Manohar:2003vb}  and NNLO in \cite{Becher:2006qw}.
The RG evolution of the jet function is given by
\begin{equation}
 \frac{dJ(p^2,\mu)}{d\ln\mu} = \left( -2 \Gamma_{\rm cusp}
\ln\frac{p^2}{\mu^2} - 2 \gamma^J \right)J(p^2,\mu)
+2\Gcusp\int^{p^2}_0
dq^2\,\frac{J(p^2,\mu)-J(q^2,\mu)}{p^2-q^2}.
\end{equation}
This integro-differential evolution equation can be solved by using the Laplace transformed jet function~\cite{Becher:2006nr,Becher:2006mr}:
\begin{equation}\label{eqs:jetfunction}
 \widetilde{j}(\ln\frac{Q^2}{\mu^2},\mu)=\int^\infty_0
dp^2\,\exp(-\frac{p^2}{Q^2e^{\gamma_E}}) J(p^2,\mu),
\end{equation}
which satisfies the the RG equation
\begin{equation}\label{eqs:jetRG}
 \frac{d}{d\ln\mu}\widetilde{j}(\ln\frac{Q^2}{\mu^2},\mu)=\left(-2
\Gamma_{\rm cusp}
\ln\frac{Q^2}{\mu^2}-2\gamma^J\right)\widetilde{j}(\ln\frac{Q^2}{\mu^2},\mu).
\end{equation}
Then the jet function at an arbitrary scale $\mu$ is given
by
\begin{equation}
 {J}(p^2,\mu)=\exp \bigl[ -4S(\mu_j,\mu)+2a^J(\mu_j,\mu)
\bigr] \widetilde{j}(\partial_{\eta_j}, \mu_j )  \frac{1}{p^2} \left(
\frac{p^2}{\mu^2_j}\right)^{\eta_j}
\frac{e^{-\gamma_E
\eta_j}}{\Gamma(\eta_j)},
\label{jet function}
\end{equation}
where $\eta_j=2 a_\Gamma(\muj,\mu)$.
The $\mu$-dependent part of the Laplace transformed jet function $\widetilde{j}(L,\mu)$ is determined by the anomalous dimensions of the jet function
as in Eq. (\ref{eqs:jetRG}), while the $\mu$-independent part can only be obtained by a fixed-order calculation.
At NLO, it is
\begin{eqnarray}\label{jet function_nlo}
\widetilde{j}(L,\mu)&=&1+\frac{\alpha_s}{4\pi}\bigg( \frac{\Gamma_0}{2}L^2+\gamma^J_0L+c^J_1\bigg),
\end{eqnarray}
with $c_1^J =\left( 7- \frac{2}{3}\pi^2 \right)C_F$.

\subsection{Soft function}
\begin{figure}
  \includegraphics[width=0.8\linewidth]{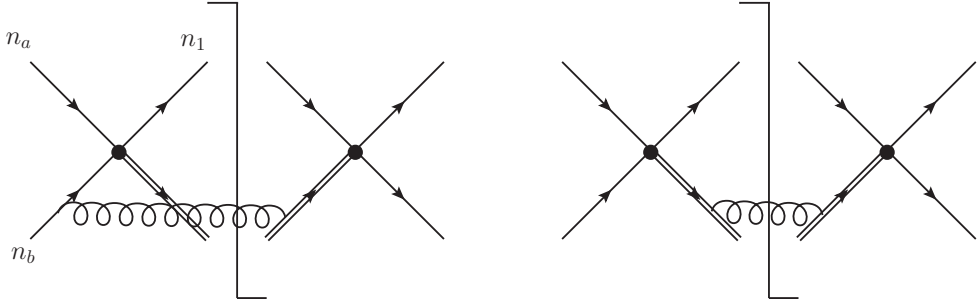}\\
  \caption{Nonvanishing diagrams contributing to the soft function at NLO.
  The contributions from the left and right diagrams are denoted as $S^{(1)}_{bt}$ and $S^{(1)}_{tt}$, respectively.}
  \label{fig:soft function}
\end{figure}

The soft function $\mathcal{S}(k^+,\mu)$, defined in Eq. (\ref{eqs:sft}), describes soft interactions between all colored particles.
It can be calculated in SCET or in the full theory in
the eikonal approximation. The LO soft function is given in Eq. (\ref{eqs:softLO}).
At NLO, we only need to calculate the nonvanishing real emission diagrams in dimensional regularization,
as shown in Fig.~\ref{fig:soft function},
which give
\begin{eqnarray}
S^{(1)}_{bt}(k,\mu)&=&\frac{2g_s^2C_F\mu^{2\epsilon}}{(2\pi)^{d-1}}\half
\int_0^{\infty}\hspace{-0.2cm}\mathrm{d}q^+\hspace{-0.1cm}
\int_0^{\infty}\hspace{-0.2cm}\mathrm{d}q^-\hspace{-0.1cm}\int\mathrm{d}^{d-2}q_{\bot}
\delta(q^+ q^--q_{\bot}^2)\delta(k-n_1\cdot q)\frac{n_b\cdot
v}{(q\cdot n_b)(q\cdot v)},
\label{eqs:soft}
\end{eqnarray}
and
\begin{eqnarray}
S^{(1)}_{tt}(k,\mu)&=&\frac{-g_s^2C_F\mu^{2\epsilon}}{(2\pi)^{d-1}}\half
\int_0^{\infty}\hspace{-0.2cm}\mathrm{d}q^+\hspace{-0.1cm}
\int_0^{\infty}\hspace{-0.2cm}\mathrm{d}q^-\hspace{-0.1cm}\int\mathrm{d}^{d-2}q_{\bot}
\delta(q^+ q^--q_{\bot}^2)\delta(k-n_1\cdot q)\frac{1}{(q\cdot
v)^2},
\end{eqnarray}
respectively. After calculating these integrals by the approach of
Ref.~\cite{Becher:2009th}, we get
\begin{eqnarray}
S_{bt}^{(1)}(k,\mu)&=&\frac{2C_F\alpha_s}{4\pi}\biggl\{
 4\biggl[ \frac{\ln \frac{k}{\tilde{\mu}}}{k} \biggr]_{\star}^{[k,\tilde{\mu}]}+\delta(k)c_{bt}^S \biggr\},\nn
 \end{eqnarray}
and
\begin{eqnarray}
S_{tt}^{(1)}(k,\mu)&=&\frac{2C_F\alpha_s}{4\pi}\biggl\{ -\biggl[
\frac{2}{k} \biggr]_{\star}^{[k,\tilde{\mu}]}+\delta(k)c_{tt}^S
\biggr\},
\end{eqnarray}
respectively, where
$\tilde{\mu}=\mu/\sqrt{(2n_{b\bar{b}})/n_1^{+2}}=(\mu(-\hat{u})m_t)/(2(-\hat{t}+m_t^2)E_1)$.
The detail of our calculations and  explicit expressions of $c_{bt}^S$ and $c_{tt}^S$ are given in
Appendix \ref{sec:soft}.
The star distribution is defined as \cite{Schwartz:2007ib}
\begin{eqnarray}
  \left[f(x)\right]_{\star}^{[x,a]} &=& f(x) {\rm ~for~} x>0 \\
  \int_0^a dx \left[f(x)\right]_{\star}^{[x,a]} g(x) &=& \int_0^a dx f(x)\left[g(x)-g(0)\right].
\end{eqnarray}
And the soft function
$\mathcal{S}(k,\mu)=S_{bt}(k,\mu)+S_{tt}(k,\mu)$, similar to the jet function,
satisfies the RG equation
\begin{eqnarray}
\frac{d }{d \ln \mu}\mathcal{S}(k,\mu)=\biggl[ -2\Gamma_{\rm cusp}\ln\frac{k}{\tilde{\mu}}+2\gamma^S \biggr]\mathcal{S}(k,\mu)
+2\Gamma_{\rm cusp}\int_0^k d k^{\prime}\frac{\mathcal{S}(k,\mu)-\mathcal{S}(k^{\prime},\mu)}{k-k^{\prime}}.
\end{eqnarray}
The solution to this equation is
\begin{eqnarray}\label{eqs:soft_function}
\mathcal{S}(k,\mu)=\exp \bigl[ -2S(\mu_s,\mu)-2a^S(\mu_s,\mu)
\bigr]\widetilde{s} (\partial_{\eta_s},\mu_s)\frac{1}{k}\biggl(
\frac{k}{\tilde{\mu}_s} \biggr)^{\eta_s}
\frac{e^{-\gamma_E\eta_s}}{\Gamma(\eta_s)},
\end{eqnarray}
where $\eta_s=2a_{\Gamma}(\mu_s,\mu)$. The Laplace transformed soft
function $\widetilde{s}(L,\mu)$ at NLO is given by
\begin{eqnarray}\label{eqs:soft_function_nlo}
\widetilde{s}(L,\mu)&=&1+\frac{\alpha_s}{4\pi}\bigg( \Gamma_0 L^2-2\gamma^S_0 L+ c^S_1 \bigg), 
\end{eqnarray}
with $c_1^S=(2c^S_{bt}+2c^S_{tt}+\frac{2\pi^2}{3})C_F$.

\subsection{Scale independence}
In the factorization formalism, we have introduced the hard function, jet function and soft function.
Each of them is evaluated at a scale to make the perturbative expansion reliable, and then evolved
to a common scale. Therefore, it is important to check the scale independence of the final results.
If we expand the exponent in Eq. (\ref{eqs:hardup}), then we can find that the dependencies on the
intermediate scale $\mu_{h,up}$ cancel each other up to $\mathcal{O}(\alpha_s)$.
The same situation happens for $\mu_{h,dn}$ in Eq. (\ref{eqs:harddn}).
The case for the jet scale is more complicated due to the partial derivative operator
and the delta function after we use the expansion
\begin{equation}\label{eqs:expan}
    \frac{1}{p^2} \left(\frac{p^2}{\mu^2_j}\right)^{\eta_j}=\frac{\delta (p^2)}{\eta_j}
    +\left[\frac{1}{p^2}\right]_{\star}^{[p^2,\mu_j^2]}
    +\eta_j\left[\frac{\ln (p^2/\mu_j^2)}{p^2}\right]_{\star}^{[p^2,\mu_j^2]}+\mathcal{O}(\eta_j^2).
\end{equation}
We point out that the scale independence happens for the jet function only in the sense of the integration over $p^2$.
The case for the soft scale is the same as for the jet scale.

After checking the intermediate scale independence, we discuss the case for the final common scale.
Recalling the hadronic threshold definition in Eq. (\ref{eqs:s4}) and the cross section near the threshold in Eq. (\ref{eqs:facmain2}),
we have
\begin{eqnarray}\label{eqs:scaleind}
\frac{d\sigma}{dS_4dy} \propto && \int dx_a dx_b\int dp_1^2 \int dk^+\frac{1}{\hat{s}}
f_{i/P_a}(x_a,\mu) f_{j/P_b}(x_b,\mu) H_{up}(\mu) H_{dn}(\mu) \nn\\&&
J(p_1^2,\mu) S(k^+,\mu) \delta(S_4-(-\hat{t})(1-x_a)-(-\hat{u})(1-x_b)-p_1^2-2k^+E_1),
\end{eqnarray}
where we have changed the integration variables $d\hat{t}d\hat{u}$ to $dp_T^2 dy$ and then to $dS_4 dy$.
From this equation, we can see clearly the connection between the threshold region of the whole system, represented by $S_4$,
and those of the parts of the system, represented by $(1-x_a),(1-x_b), p_1^2$ and $ k^+$, respectively.
To change the convolution form to a simpler product form, we apply the Laplace transformation to the above equation and obtain
\begin{equation}
    \frac{d\tilde{\sigma}}{dQ^2dy}=\int_0^{\infty} dS_4 \exp \left(-\frac{S_4}{Q^2e^{\gamma_E}}\right)\frac{d\sigma}{dS_4dy}.
\end{equation}
The Laplace transformed jet function and its RG evolution are given in Eq. (\ref{eqs:jetfunction}) and Eq. (\ref{eqs:jetRG}).
Here, for convenience, we write its RG equation again as
\begin{equation}
 \frac{d}{d\ln\mu}\widetilde{j}(\ln\frac{Q^2}{\mu^2},\mu)=\left(-2
\Gamma_{\rm cusp}
\ln\frac{Q^2}{\mu^2}-2\gamma^J\right)\widetilde{j}(\ln\frac{Q^2}{\mu^2},\mu).
\end{equation}
The Laplace transformed soft function is similar to the jet function, but its RG equation is
\begin{equation}
 \frac{d}{d\ln\mu}\widetilde{s}(\ln \frac{Q^2(-\hat{t}+m_t^2)}{\mu(-\hat{u})m_t},\mu)=\left(-2\Gamma_{\rm cusp}\ln \frac{Q^2(-\hat{t}+m_t^2)}{\mu(-\hat{u})m_t}
+2\gamma^S\right)\widetilde{s}(\ln \frac{Q^2(-\hat{t}+m_t^2)}{\mu(-\hat{u})m_t},\mu).
\end{equation}
The Laplace transformed PDF near the endpoint is given by
\begin{equation}
\tilde{f}_{i/P}(\tau,\mu)=\int_0^1 dx \exp \left(-\frac{1-x}{\tau e^{\gamma_E}}\right) f_{i/P_a}(x,\mu),
\end{equation}
which satisfies the RG equation
\begin{equation}
 \frac{d}{d\ln\mu}\tilde{f}_{i/P}(\tau,\mu)=\left(2\Gamma_{\rm cusp}
\ln (\tau) + 2\gamma^{\phi}\right)\tilde{f}_{i/P}(\tau,\mu).
\end{equation}
Due to the delta function in Eq. (\ref{eqs:scaleind}), the variables $\tau_{a,b}$ in the Laplace transformed PDF are given by
\begin{equation}
    \tau_a = \frac{Q^2}{-\hat{t}}{\rm~~~~for~~~~} \tilde{f}_{i/P_a}(\tau_a,\mu),\quad {\rm and} \quad
    \tau_b = \frac{Q^2}{-\hat{u}}{\rm~~~~for~~~~} \tilde{f}_{j/P_b}(\tau_b,\mu).
\end{equation}
For completeness, we also need the RG equations for the hard functions which have been given by Eq. (\ref{eqs:hardupRG}) and Eq. (\ref{eqs:harddnRG}).
We rewrite them as
\begin{eqnarray}
\frac{d}{d~\mathrm{ln}\mu}H_{up}(\mu)&=&\biggl(2\Gamma_{\rm cusp}~\mathrm{ln}\frac{-\hat{t}}{\mu^2}+2\gamma_{up}^V \biggr )H_{up}(\mu),\\
\frac{d}{d~\mathrm{ln}\mu}H_{dn}(\mu)&=&\biggl(2\Gamma_{\rm cusp}~\mathrm{ln}\frac{-\hat{t}+m_t^2}{\mu m_t}+2\gamma_{dn}^V \biggr )H_{dn}(\mu).
\end{eqnarray}

So far, we can check the scale independence of the final results.
Using the relation between anomalous dimensions given in Eq. (\ref{eqs:anormdim}), we can immediately obtain
\begin{equation}
    \frac{d}{d\ln \mu}\left[\tilde{f}_{i/P_a}(\tau_a,\mu) \tilde{f}_{j/P_b}(\tau_b,\mu) H_{up} (\mu) H_{dn} (\mu)
    \widetilde{j}(\ln\frac{Q^2}{\mu^2},\mu)\widetilde{s}(\ln \frac{Q^2(-\hat{t}+m_t^2)}{\mu(-\hat{u})m_t},\mu)\right]=0.
\end{equation}
Even more precisely, we have
\begin{eqnarray}
    \frac{d}{d\ln \mu}\left[\tilde{f}_{i/P_a}(\tau_a,\mu)  H_{up} (\mu)
    \widetilde{j}(\ln\frac{Q^2}{\mu^2},\mu)\right]&=&0, \nn\\
    \frac{d}{d\ln \mu}\left[ \tilde{f}_{j/P_b}(\tau_b,\mu)  H_{dn} (\mu)
    \widetilde{s}(\ln \frac{Q^2(-\hat{t}+m_t^2)}{\mu(-\hat{u})m_t},\mu)\right]&=&0.
\end{eqnarray}
This means that if we evolve the scales of $H_{up}$ and the jet function to the factorization scale of the light quark line $\mu_{F,up}$,
then the final results should not depend on $\mu_{F,up}$.
And if we evolve the scales of $H_{dn}$ and the soft function to the factorization scale of the heavy quark line $\mu_{F,dn}$,
then the final results should not depend on $\mu_{F,dn}$.
Actually, the relationships between the anomalous dimensions given in Eq. (\ref{eqs:anormdim}) are determined by these requirements.

\subsection{Final RG improved differential cross section}
After combining the hard, jet and soft functions together, we obtain the resummed differential cross
section for t-channel single top production
\begin{eqnarray}
\frac{d\hat{\sigma}^{\rm thres}}{d\hat{t}d\hat{u}} &=&
\sum_{ij}\frac{\lambda_{0,ij}}{64\pi N^2_c\hat{s}^2} \nn\\
&& \exp\bigl[ 4S(\mu_{h,up},\mu_{F,up})-2a_{up}^V(\mu_{h,up},\mu_{F,up})\bigr]
  \bigg( \frac{-\hat{t}}{\mu_{h,up}^2} \bigg)^{-2a_{\Gamma}(\mu_{h,up},\mu_{F,up})}H_{up}(\mu_{h,up}) \nn\\
&& \exp\bigl[ 2S(\mu_{h,dn},\mu_{F,dn})-2a_{dn}^V(\mu_{h,dn},\mu_{F,dn})\bigr]
  \bigg( \frac{-\hat{t}+m_t^2}{\mu_{h,dn}m_t} \bigg)^{-2a_{\Gamma}(\mu_{h,dn},\mu_{F,dn})}H_{dn}(\mu_{h,dn})\nn\\
&& \exp \bigl[ -4S(\mu_j,\mu_{F,up})+2a^J(\mu_j,\mu_{F,up})\bigr] \left( \frac{m_t^2}{\mu^2_j} \right)^{\eta_j} \nn\\
&& \exp \bigl[ -2S(\mu_s,\mu_{F,dn})-2a^S(\mu_s,\mu_{F,dn}) \bigr]\biggl( \frac{m_t(-\hat{t}+m_t^2)}{\mu_s(-\hat{u}) } \biggr)^{\eta_s}  \nn\\
&& \widetilde{j}(\partial_{\eta}+L_j , \mu_j )\widetilde{s}(\partial_{\eta}+L_s,\mu_s)
 \frac{1}{s_4}\left( \frac{s_4}{m_t^2} \right)^{\eta}
 \frac{e^{-\gamma_E\eta}}{\Gamma(\eta)},
\end{eqnarray}
where $\eta=\eta_j+\eta_s$ and $L_j=\ln (m_t^2/\mu^2_j), L_s=\ln [m_t(-\hat{t}+m_t^2)/\mu_s(-\hat{u})]$.
In the above expression, the hard function $H_{up}$ and jet function ($H_{dn}$ and soft function) have been evolved to the scale $\mu_{F,up}$ ($\mu_{F,dn}$).
It seems that the t-channel single top production is factorized as two DIS processes.
However, the convolution of the jet and soft functions, now expressed in terms of the partial derivative operator acting on the same kernel function,
violates this simple factorization and connects the two DIS processes nontrivially.

If we set scales $\mu_{h,up}$, $\mu_{h,dn}$, $\mu_{j}$, $\mu_{s}$
equal to the common scale $\mu$, which is conveniently chosen as the
factorization scale $\mu_{F,up}=\mu_{F,dn}=\mu_{F}$, then we recover
the threshold singular plus distributions, which should appear in
the fixed-order calculation. Up to order $\alpha_s^2$, we have
\begin{eqnarray}\label{eqs:singular}
\left(\frac{\lambda_{0,ij}}{64\pi N^2_c\hat{s}^2}\right)^{-1}\frac{d\hat{\sigma}_{ij}^{\rm thres}}{d\hat{t}d\hat{u}} &=&\delta(s_4)
+\frac{\alpha_s}{4\pi}\biggl\{A_2 D_2+A_1 D_1 + A_0 \delta(s_4) \biggr\} \nn\\
&&\hspace{-3.0cm}+\bigg(\frac{\alpha_s}{4\pi}\bigg)^2 \bigg\{ B_4 D_4+ B_3 D_3+ B_2 D_2+ B_1 D_1 + B_0 \delta(s_4)\bigg\},
\end{eqnarray}
where
\begin{equation}\label{eqs:D}
    D_n=\biggr[\frac{\ln^{n-1} (s_4/m_t^2)}{s_4} \biggl]_{+}.
\end{equation}
The $A_n$ and $B_n$ coefficients are given by
\begin{eqnarray}
  A_2 &=& 3\Gamma_0, \\
  A_1 &=& (L_j+2L_s)\Gamma_0 + \gamma^J_0-2\gamma^S_0, \\
  A_0 &=& \left(-\frac{1}{2} L_{h,up}^2- L_{h,dn}^2 + \frac{1}{2} L_j^2 + L_s^2 - \frac{\pi^2}{4}\right)\Gamma_0
          - \gamma^V_{up,0} L_{h,up}  - 2\gamma^V_{dn,0} L_{h,dn} + \gamma^J_0 L_j - 2\gamma^S_0 L_s \nn\\&&
          + c^H_1+ c^J_1+ c^S_1,\\
  B_4 &=& \frac{A_2^2}{2}, \\
  B_3 &=& \frac{9}{2}A_1\Gamma_0-\frac{5}{2}\beta_0\Gamma_0, \\
  B_2 &=& A_1^2+A_2A_0-\frac{3\pi^2}{2}\Gamma_0^2 -\beta_0\Gamma_0(L_j+4L_s) -\beta_0(\gamma^J_0-4\gamma^S_0)+3\Gamma_1 ,\\
B_1 &=& A_1\left(A_0-\frac{\pi^2}{2}\Gamma_0\right)+9\zeta_3\Gamma^2_0
-\beta_0\Gamma_0\left(\half L_j^2 + 2L_s^2-\frac{5}{12}\pi^2\right)-\beta_0(c^J_1+2c^S_1+\gamma^J_0L_j-4\gamma^S_0L_s)\nn\\&&
+\gamma^J_1-2\gamma^S_1+\Gamma_1(L_j+2L_s)
,
\end{eqnarray}
where $\zeta_3=1.20206\cdots$, $L_{h,dn}=\ln [(-\hat{t}+m_t^2)/\mu m_t]$, $L_{h,up}=\ln (-\hat{t}/\mu^2)$ and $c^H_1=c^{H,up}_1+c^{H,dn}_1$.
We find that $A_{2,1}$, $B_{4,3,2}$ and the scale-dependent parts of $A_0$ and $B_1$ agree with the results in Ref.~\cite{Kidonakis:2011wy}
with the replacement $\hat{t}(\hat{u})\to \hat{t}(\hat{u})-m_t^2$ due to the different definition of $s_4$ there.

To give  precise predictions, we resum the singular terms to all orders and include the nonsingular terms up to NLO.
We obtain the final RG improved differential cross section
\begin{eqnarray}
\frac{d\hat{\sigma}_{\rm Resum}}{d\hat{t}d\hat{u}}&=&\frac{d\hat{\sigma}^{\rm thres}}{d\hat{t}d\hat{u}}
  +\left( \frac{d\hat{\sigma}_{\rm NLO}}{d\hat{t}d\hat{u}} -\frac{d\hat{\sigma}^{\rm thres}}{d\hat{t}d\hat{u}}\right)\Big |_{\rm expanded~to~NLO}.
\end{eqnarray}
Near the threshold regions, the expansion of the resummed result approaches the fixed-order one so that the second term in the above equation almost vanishes and
the threshold contribution dominates.
In the regions far from the threshold limit, the resummation effect is not important and the final result is mainly determined by the fixed-order calculations.

\section{Numerical Discussion}
\label{sec:nume}
In this section, we discuss the numerical results for threshold
resummation effects on t-channel single top production at the Tevatron ($\sqrt{s}=1.96$ TeV) and the LHC ($\sqrt{s}=8$ TeV).
The top quark mass is chosen as $m_t=173.2 {\rm ~ GeV}$ \cite{Lancaster:2011wr} and the rapidity is integrated over $-2.4<y<2.4$ if not specified explicitly.
For the $W$ boson mass we take $M_W=80.4$ GeV.
We set the Fermi constant to be $G_F=1.1664\times 10^{-5}{\rm ~ GeV^{-2}}$.
The CKM matrix is given by
\begin{equation}
V_{CKM}=
\left(
  \begin{array}{ccc}
    0.9751 & 0.2215 & 0.0035 \\
    0.2210 & 0.9743 & 0.0410 \\
    0 & 0 & 1 \\
  \end{array}
\right)
\end{equation}
Throughout the numerical calculations, we use the MSTW2008nnlo PDF sets and associated strong coupling constant.
The factorization scales are set at $m_t$ unless specified otherwise.
There are four other scales, i.e., $\mu_{h,up},\mu_{h,dn},\mu_j,\mu_s$, introduced in the factorization formalism.
They should be properly chosen so that the corresponding hard functions, jet function and soft function have stable perturbative expansions.
In order to achieve this aim, each function should not contain large logarithms.
From Eqs. (\ref{hard_function})-(\ref{hardf_function}), we can see that if we choose $\mu_{h,up}=Q=\sqrt{-\hat{t}}$ and $\mu_{h,dn}=(Q^2+m_t^2)/m_t$,
then the large logarithms disappear.
Also as discussed below Eq. (\ref{eqs:hardmatrix}), if we combine the two hard functions blindly,
we cannot choose a proper hard scale to eliminate all the large logarithms simultaneously.
This is due to the fact that intrinsically the $W$ boson connects interactions at different scales.
\begin{figure}
  \includegraphics[width=0.3\linewidth]{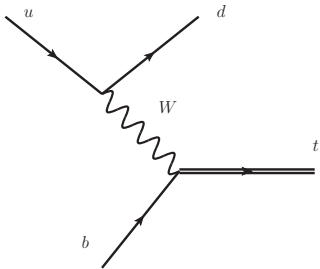}\\
  \caption{The Feynman diagrams for the single top production via the fusion of a $W$ boson and a bottom quark.}
  \label{fig:lo-fusion}
\end{figure}
We can take another viewpoint on the t-channel single top production and consider it to be a fusion process, as shown in Fig. \ref{fig:lo-fusion}.
An initial state up quark emits a $W$ boson, which then combines with a bottom quark to produce a single top quark.
The production of the $W$ boson is similar to a DIS process and there is no specific constraint on the virtuality of the $W$ boson.
But when it coannihilates with a bottom quark, the mass of final state top quark impose constraints on the `initial' $W$ boson.
As a result, the typical scales of the interactions involving the light quarks and top quarks are $Q$ and $m_t$, respectively,
which are just about the natural hard scales.

\begin{figure}
  \includegraphics[width=0.4\linewidth]{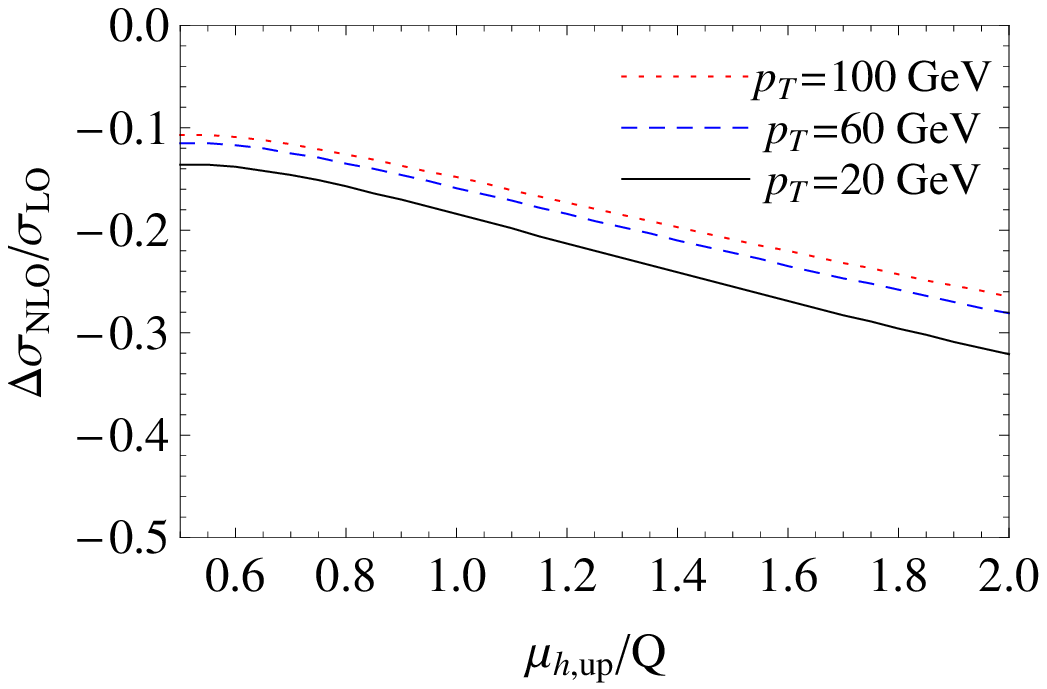}
  \includegraphics[width=0.4\linewidth]{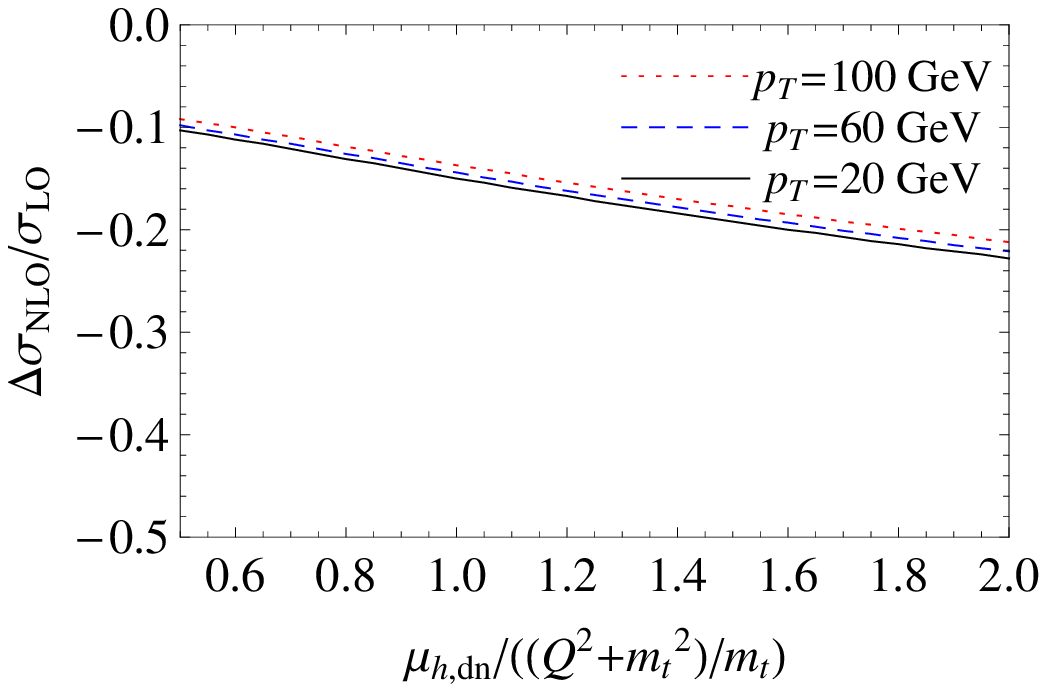}\\
  \caption{The contributions from hard functions to the NLO cross section.}
  \label{fig:muh}
\end{figure}

For the jet and soft scales, the situations are not so clear.
After inspection of Eqs. (\ref{jet function})-(\ref{jet function_nlo}) and (\ref{eqs:soft_function})-(\ref{eqs:soft_function_nlo}),
one finds that the natural jet and soft scales should be $\sqrt{p^2}$ and $2kE_1(-\hat{t}+m_t^2)/(-\hat{u})/m_t$, respectively.
But these two scales are not directly connect to the integration variables in Eq. (\ref{eqs:main}).
Moreover, they can become so small that the strong coupling constants in the jet and soft functions would diverge.
Therefore, in practice, we choose the natural jet and soft scales numerically.
\begin{figure}
  \includegraphics[width=0.4\linewidth]{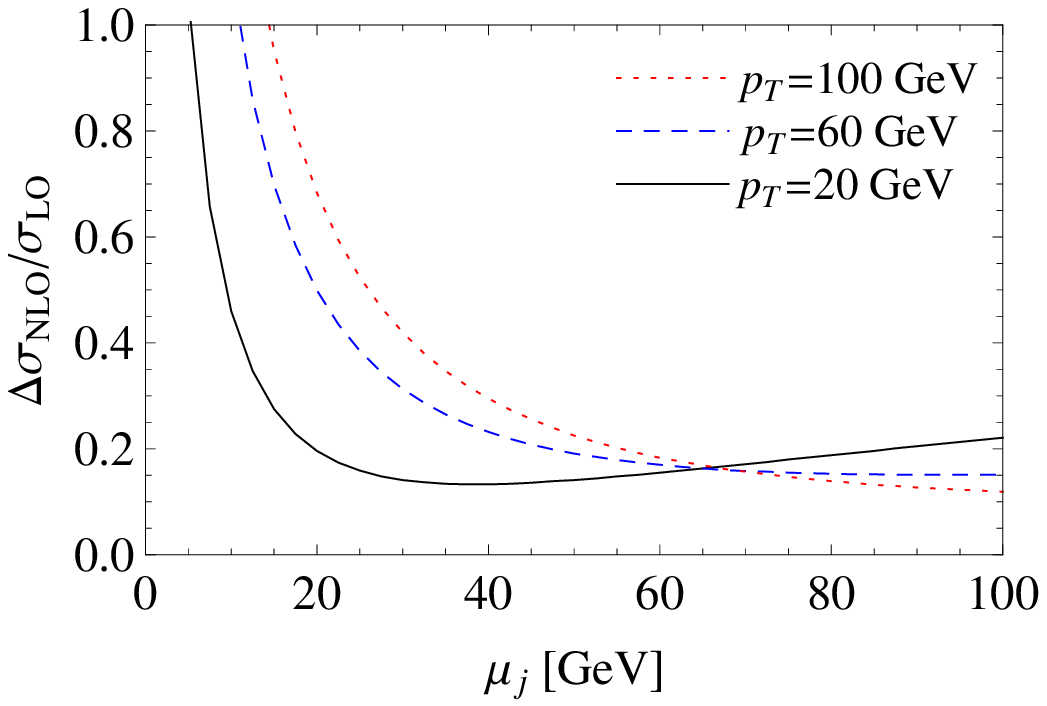}
  \includegraphics[width=0.4\linewidth]{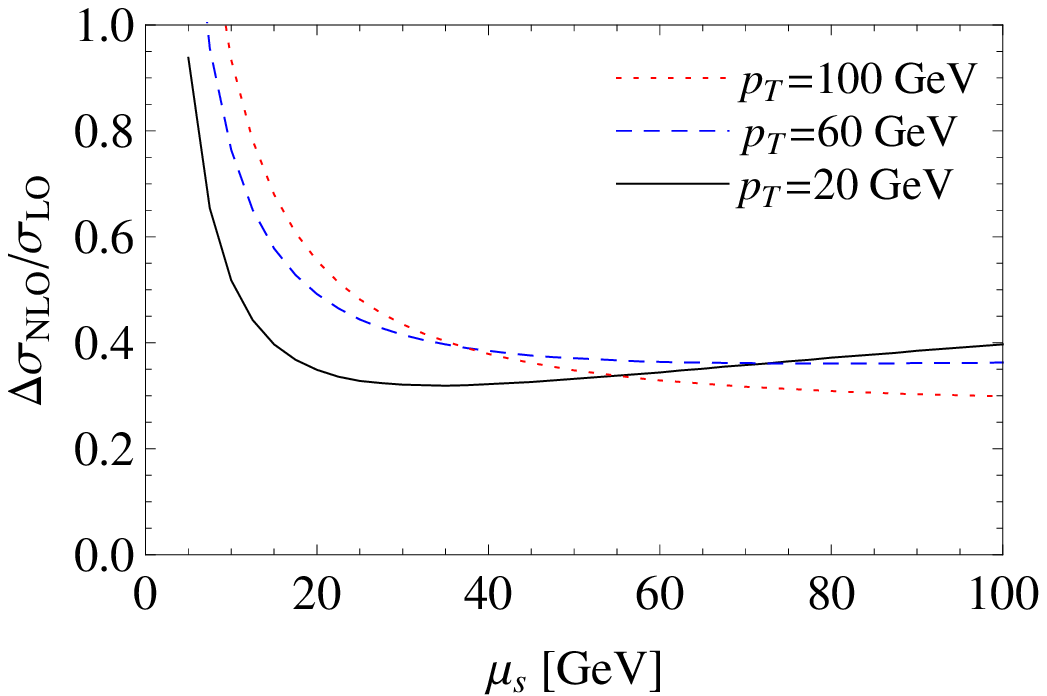}\\
  \caption{The contributions from jet and soft functions to the NLO cross section.}
  \label{fig:muj}
\end{figure}
In Fig. \ref{fig:muj}, we show the contributions to the NLO cross section from jet and soft functions separately
without including the RG evolution effects.
We have fixed the top quark transverse momentum to be $p_T=20,60,100$ GeV and change the jet (soft) scale from 5 GeV to 100 GeV.
It is required that the perturbative expansions of the jet and soft functions converge fast.
Thus, we choose the jet and soft scales as 80 GeV and 50 GeV, respectively.
When giving the final RG improved cross sections, we will investigate the scale uncertainties due to these choices.
From Fig. \ref{fig:muj}, we can also see that the jet and soft functions give positive contributions to the NLO cross sections,
and can be as large as about $20 \%$ and $40 \%$.
To see the corrections from hard functions, in Fig. \ref{fig:muh}, we show the contributions to the NLO cross section from hard functions.
We find that the hard functions provide negative contributions to the NLO cross sections and the corrections are about $-15\%$
for $\mu_{h,up}=Q$ and $\mu_{h,dn}=(Q^2+m_t^2)/m_t$.

\begin{figure}
  \includegraphics[width=0.45\linewidth]{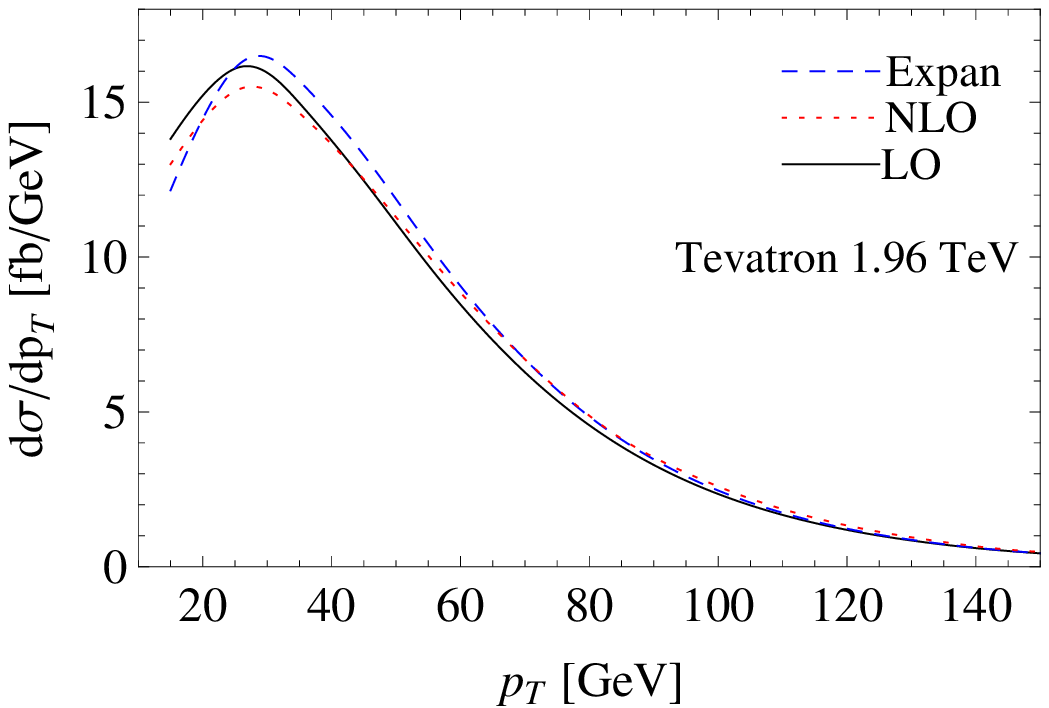} \includegraphics[width=0.45\linewidth]{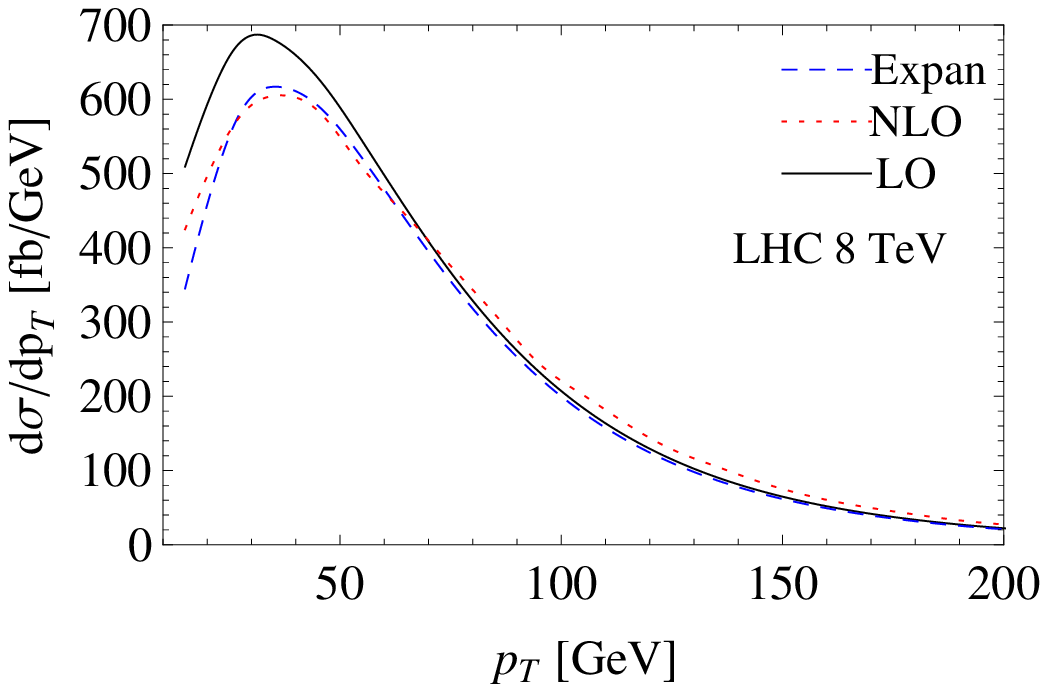}\\
  \caption{The singular terms contribution and fixed-order cross sections for t-channel single top production at the Tevatron (left) and the LHC (right).
  The dashed line represents the contributions from the singular terms up to $\mathcal{O}(\alpha_s$) which is given in Eq. (\ref{eqs:singular}).}
  \label{fig:approx}
\end{figure}

Before presenting the numerical results for the RG improved cross section, it is important to
examine to what extent the singular terms approximate the fixed-order calculation.
In Fig. \ref{fig:approx}, we present the singular terms contribution and fixed-order cross sections.
We see that the NLO cross section is well approximated by the singular terms when the top quark transverse momentum $p_T$ is larger than 50 (70) GeV at the Tevatron (LHC).
Therefore, the singular terms should be resummed for the large $p_T$ region. For the small $p_T$ region, the singular terms do not dominate the NLO corrections,
so there is no need to perform resummation in this region.
In the following discussion, we will only present the resummation results for $p_T>$ 50 (70) GeV at the Tevatron (LHC).
Meanwhile we find that the NLO QCD correction is small for t-channel single top production.
This is because the large positive soft and jet functions cancel with the large negative hard functions, as discussed in the last paragraph.
If these large effects are resummed to higher orders, we can see whether there is still a cancellation between them.

\begin{figure}
  \includegraphics[width=0.45\linewidth]{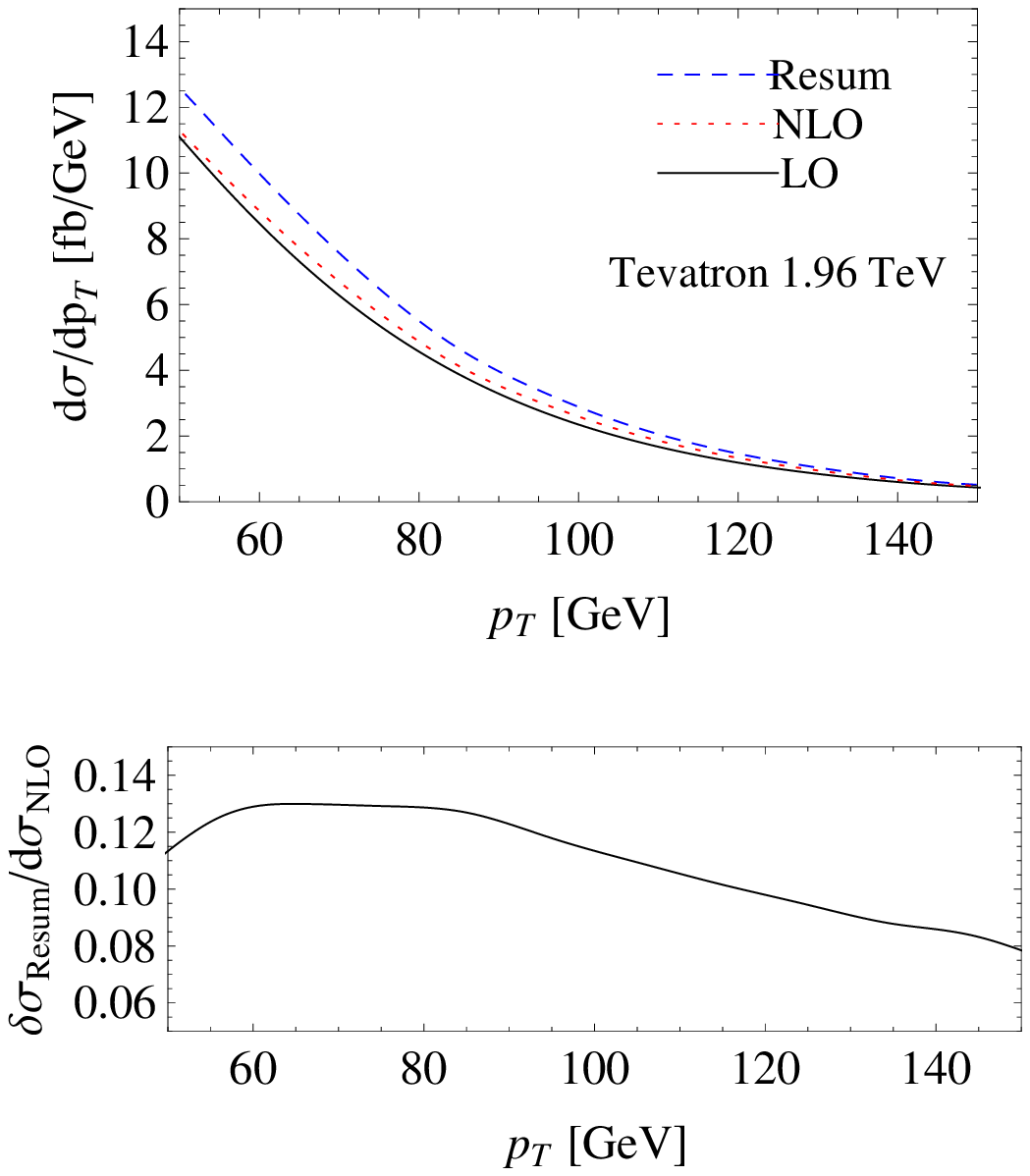} \includegraphics[width=0.45\linewidth]{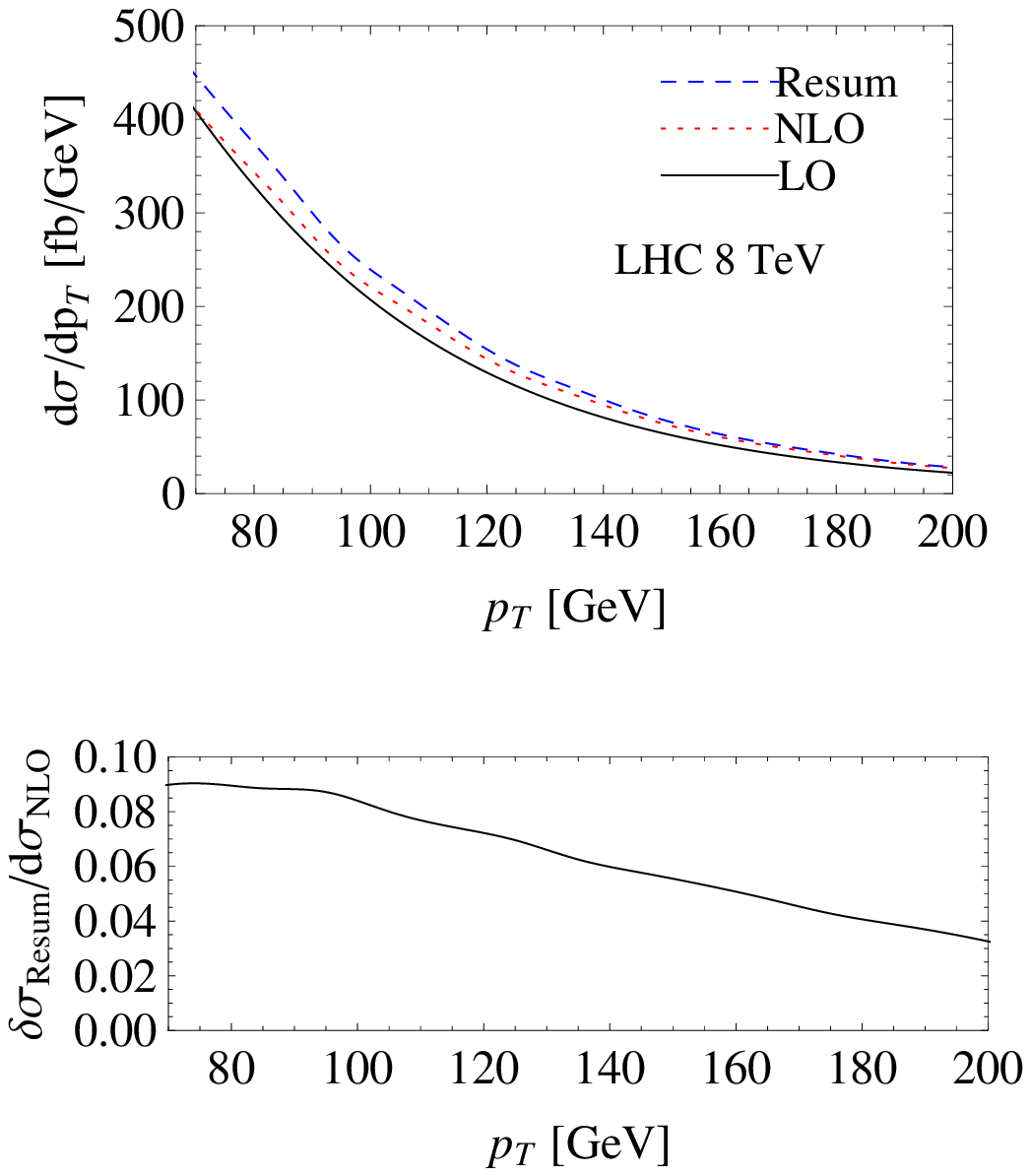}\\
  \caption{The RG improved (dashed) and fixed-order cross sections for t-channel single top production at the Tevatron (left) and the LHC (right).
  We have defined $\delta \sigma_{\rm Resum}=d\sigma_{\rm Resum}-d\sigma_{\rm NLO}$.}
  \label{fig:resum}
\end{figure}

\begin{figure}
  \includegraphics[width=0.4\linewidth]{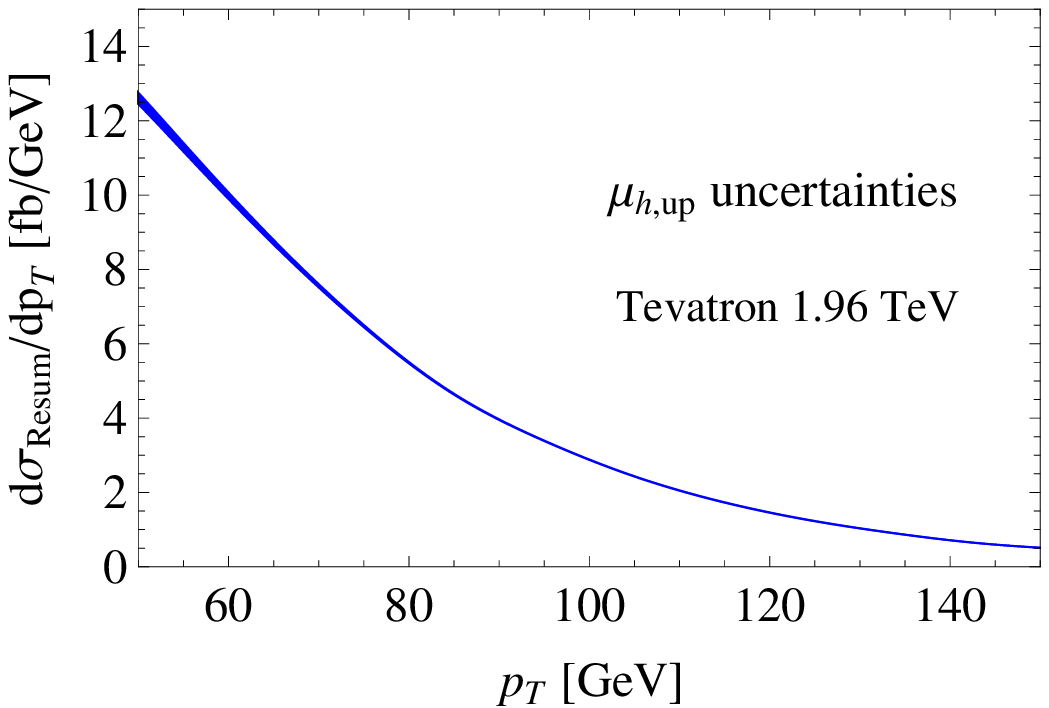}\includegraphics[width=0.4\linewidth]{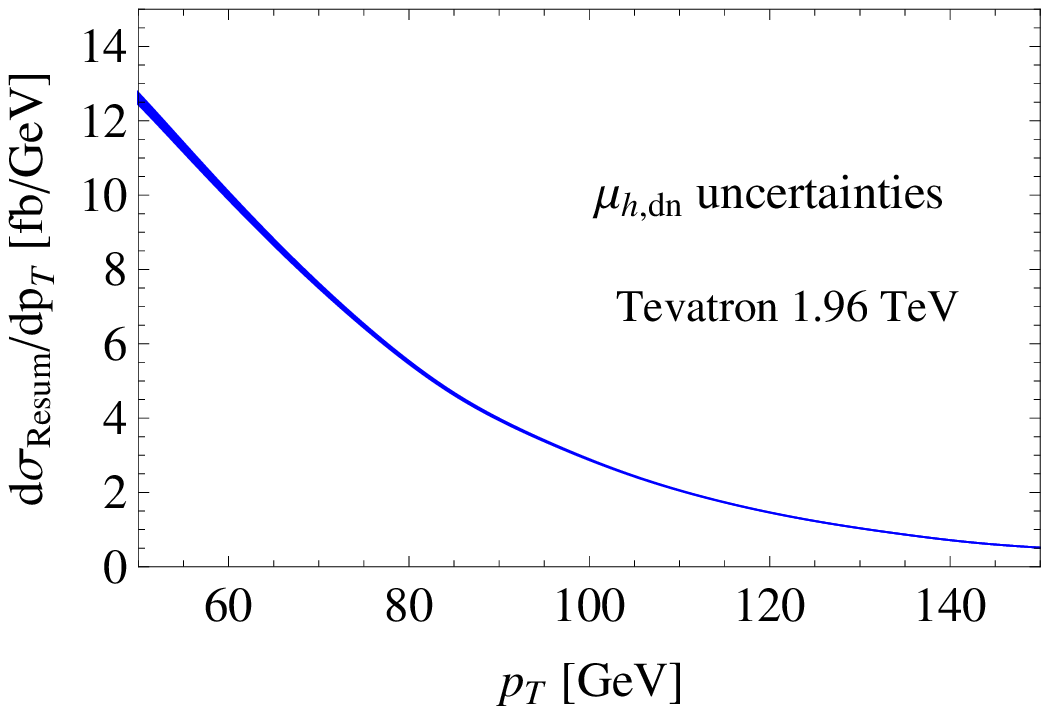}\\
  \includegraphics[width=0.4\linewidth]{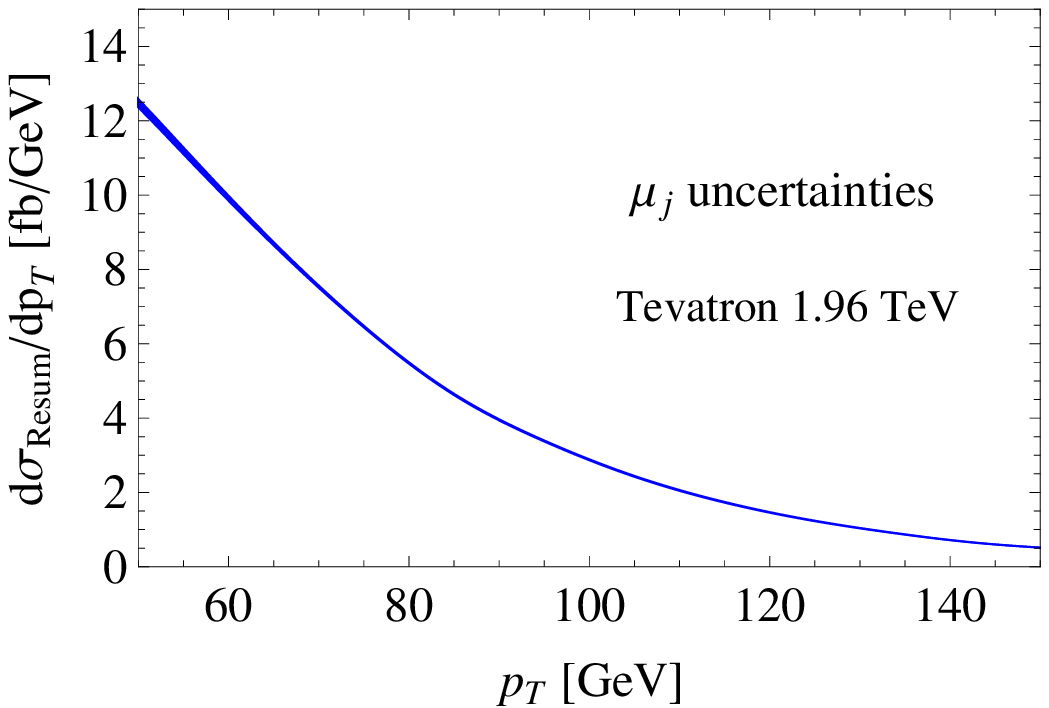}\includegraphics[width=0.4\linewidth]{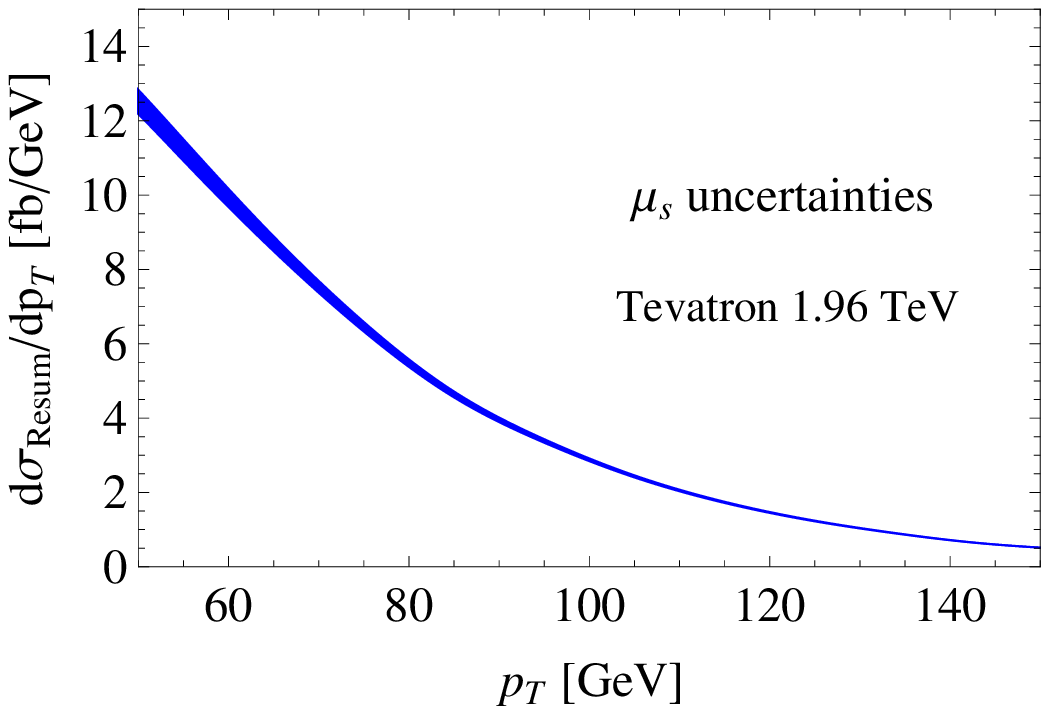}\\
  \caption{The scale uncertainties  of the resummation results due to the variations of $\mu_{h,up}$, $\mu_{h,dn}$, $\mu_{j}$, $\mu_{s}$, respectively.}
  \label{fig:uncertainties-muh}
\end{figure}

\begin{figure}
  \includegraphics[width=0.45\linewidth]{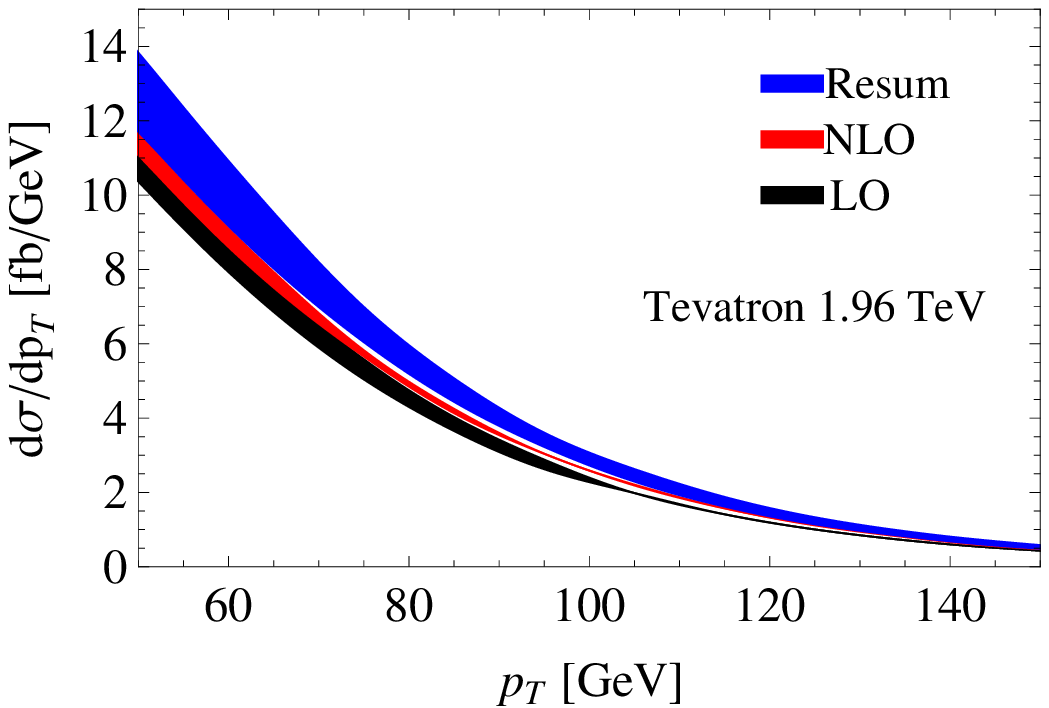}
  \includegraphics[width=0.45\linewidth]{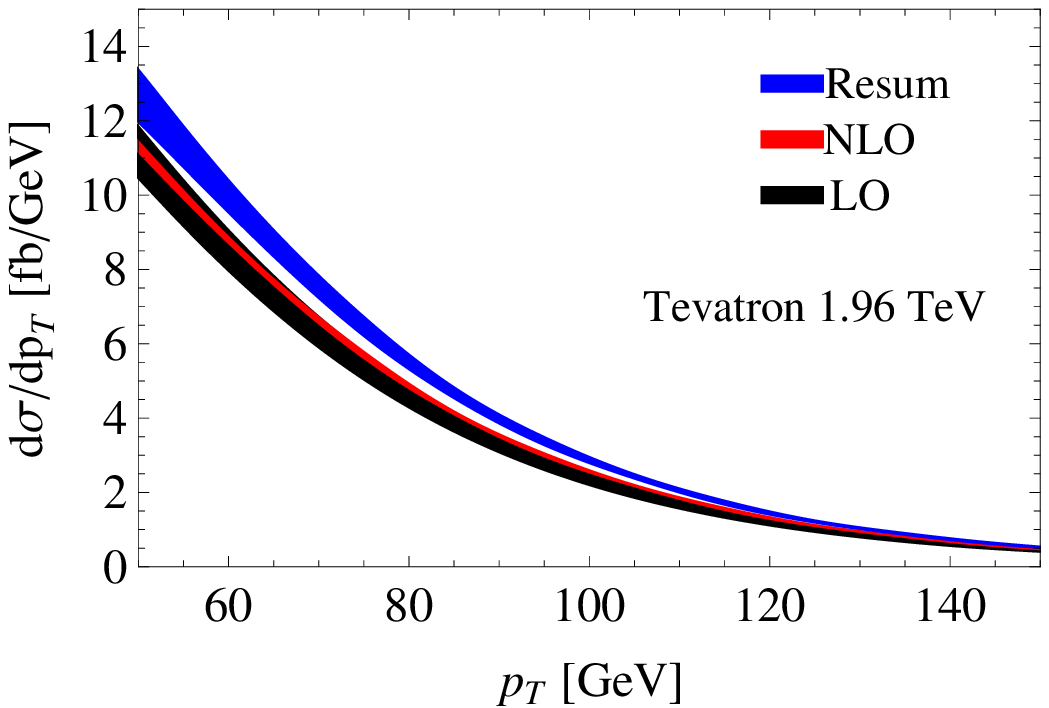}\\
  \includegraphics[width=0.45\linewidth]{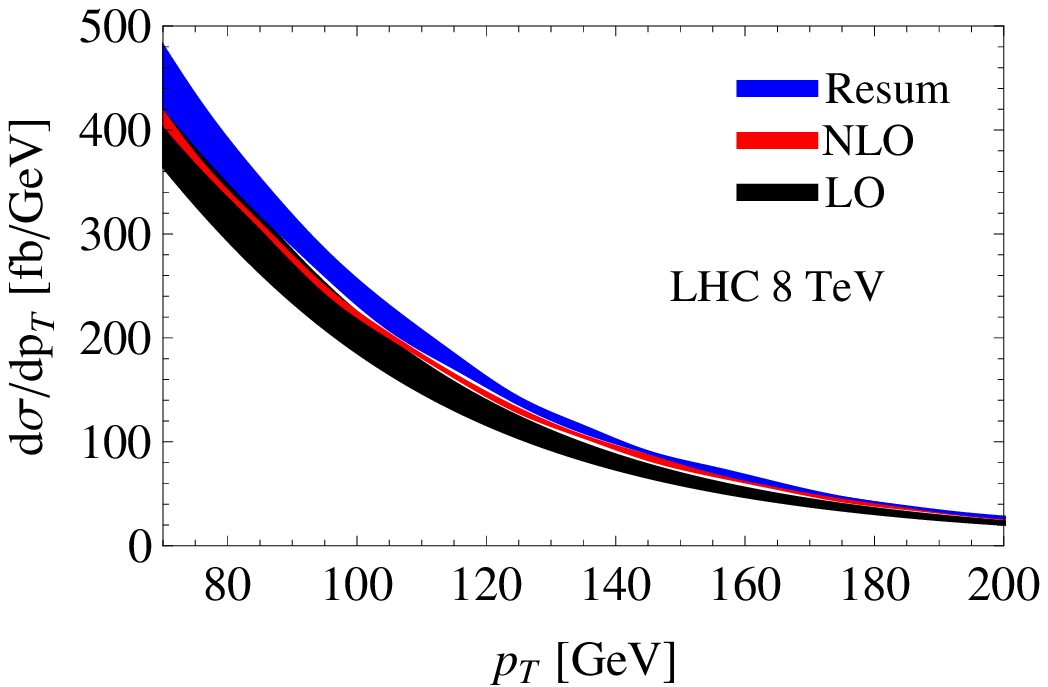}
  \includegraphics[width=0.45\linewidth]{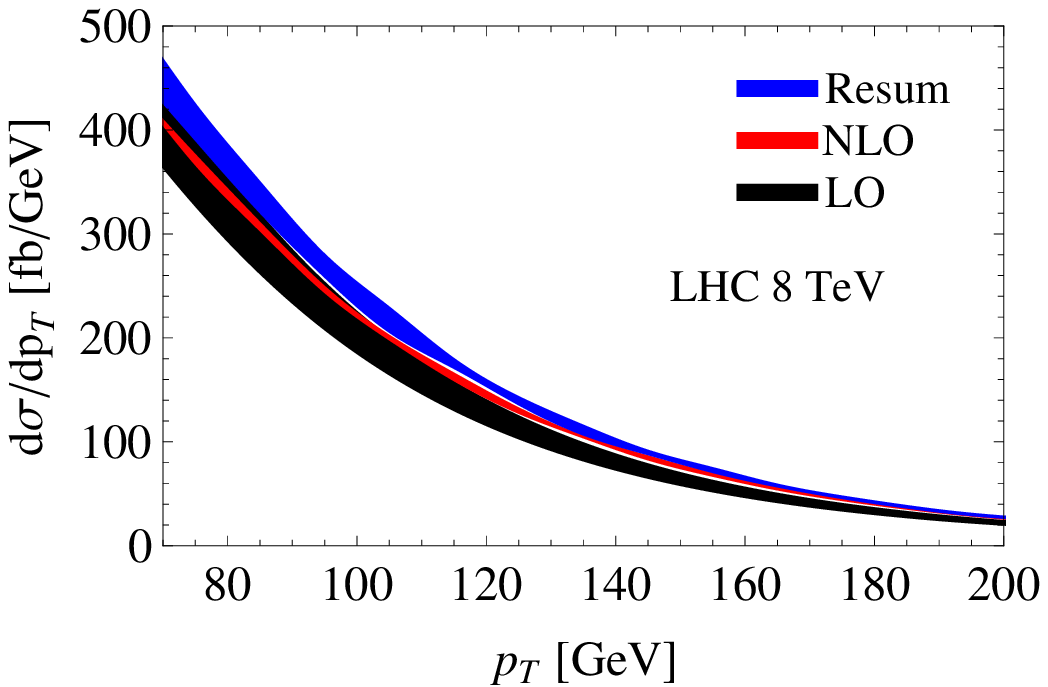}\\
  \caption{The scale uncertainties due to the variations of $\mu_{F,dn}$ (left) and $\mu_{F,up}$ (right), respectively.
  The bands in each plots from top to bottom denote the resummation, NLO and LO results, respectively.}
  \label{fig:uncertainties}
\end{figure}

We have chosen all the natural scales involved in this process. Now we
give the numerical results of the resummed cross section. When
discussing each scale dependence, we fix the other scales at the
natural scales discussed above.
In Fig. \ref{fig:resum}, we show the RG improved cross sections as a function of the top quark $p_T$.
We can see  that the distribution is increased by about $9\sim 13 \%$ and $4\sim 9 \%$ for $p_T>$ 50 and 70 GeV at the Tevatron and LHC, respectively, compared to the NLO results.
In Fig. \ref{fig:uncertainties-muh}, we give the uncertainties of the resummation results  due to the change of intermediate scales
$\mu_{h,up}$, $\mu_{h,dn}$, $\mu_{j}$, $\mu_{s}$ independently by a factor of two.
The uncertainties arising from $\mu_{h,up}$, $\mu_{h,dn}$ and $\mu_{j}$ are less than $\pm 1\%$,
and for $\mu_{s}$ are about $\pm 2\%$.
In Fig. \ref{fig:uncertainties}, we show the scale uncertainties of the resummation results due to the variations of $\mu_{F,dn}$ and $\mu_{F,up}$ by a factor of two,
and do not see scale uncertainties are decreased, compared to the NLO results. In principle, the scale uncertainties should vanish,
as illustrated analytically in the last section. However, the analysis there is based on the assumption that
the PDF is evaluated near the endpoint. But in practice, this is not always true because the center-of-mass energy
of the Tevatron or LHC is much larger than the invariant mass of the final states. And
the dynamical enhancement mechanism \cite{Becher:2007ty} is not appropriate for a t-channel process.
On the other hand, when approaching the threshold region, i.e., with the increasing of  the top quark $p_T$, the scale uncertainties of the resummed cross sections are significantly reduced, as shown in Fig. \ref{fig:uncertainties}.

\section{Conclusion}
\label{sec:conc}
We have studied the factorization and resummation of t-channel top quark transverse momentum distribution at large $p_T$
in the SM at both the Tevatron and the LHC with SCET.
This is the first spacelike process studied in SCET involving one massless and one massive colored particles in the final states.
The cross section in the threshold region can be factorized into a convolution of hard, jet and soft functions.
In particular, we first calculate the NLO soft functions for this process, and give a RG improved cross section by evolving the different functions to a common scale.
Our results show that the resummation effects increase the NLO results by about $9\%\sim 13\%$ and  $4\%\sim 9\%$
when the transverse momentum of the top quark is larger than $50$ and 70~GeV at the Tevatron  and the 8 TeV LHC, respectively.
Our prediction on the transverse momentum distribution of the top quark in the large $p_T$ region is important in the search for new physics, e.g., a heavy $W'$ which
can mediate the single top production through the s-channel.
Also, we discuss the scale independence of the cross section analytically and show how to choose the proper scales at which the perturbative expansion can converge fast.

\acknowledgments
This work was supported in part by the National Natural
Science Foundation of China, under Grants No. 11021092, No. 10975004 and No. 11135003.

\appendix

\section{Calculation of the soft functions}\label{sec:soft}
In this appendix, we present the details of the calculation of the
two $\mathcal{O}(\alpha_s)$ soft functions $S^{(1)}_{bt}(k,\mu)$ and
$S^{(1)}_{tt}(k,\mu)$. We choose to do the calculation in the rest
frame of the top quark, in which the  four-velocity of the top quark
is $v^{\mu}=(1,0,0,0)$. This choice of frame makes the denominators
simple but leaves the complexity in the delta functions.
Actually, we also perform the calculation in the frame where the delta
functions are simple but the singularities in the denominators are hard
to isolate~\cite{Kelley:2010qs}. And finally we find the same results,
which can be considered as a strong cross check for our calculations.

In the rest frame of the top quark, we also choose $n_b^{\mu}=(1,0,0,1)$. Then,
\begin{eqnarray}
q^{\mu}=q^+\frac{\bar{n}_b^{\mu}}{n_{b\bar{b}}}+q^-\frac{n_b^{\mu}}{n_{b\bar{b}}}+q_{\perp}^{\mu},\quad
n_1^{\mu}=n_1^+\frac{\bar{n}_b^{\mu}}{n_{b\bar{b}}}+n_1^{-}\frac{n_b^{\mu}}{n_{b\bar{b}}}+n_{1\perp}^{\mu},
\end{eqnarray}
and
\begin{eqnarray}
q\cdot n_1=\frac{q^+n_1^-+q^-n_1^+}{n_{b\bar{b}}}-|q_{\perp}||n_{1\perp}|\cos\theta,\quad
q\cdot v  =q \cdot \frac{(n_b+n_{\bar{b}})}{2}=\frac{(q^++q^-)}{2}.
\end{eqnarray}
After putting these expressions into the formula (\ref{eqs:soft}), we get
\begin{eqnarray}
S^{(1)}_{bt}(k,\mu)&=&\frac{g_s^2C_F\mu^{2\epsilon}}{(2\pi)^{d-1}} \int_0^{\infty}\hspace{-0.2cm}\mathrm{d}q^+\hspace{-0.1cm} \int_0^{\infty}\hspace{-0.2cm}\mathrm{d}q^-\hspace{-0.1cm}\int\mathrm{d}\Omega_{d-2}\left(\frac{2q^+q^-}{n_{b\bar{b}}} \right)^{-\epsilon}\nn\\
&&\delta(k-\frac{q^+n_1^-+q^-n_1^+}{n_{b\bar{b}}}+|q_{\perp}||n_{1\perp}|\cos\theta)\frac{n_b\cdot v}{q^+(q^++q^-)}.
\end{eqnarray}
Now redefine the integration variables $q^+$ and  $q^-$ and let $a=\frac{n_1^+}{n_1^-}$, then
\begin{eqnarray}
S^{(1)}_{bt}(k,\mu)&=&\frac{g_s^2C_F\mu^{2\epsilon}}{(2\pi)^{d-1}} \int_0^{\infty}\hspace{-0.2cm}\mathrm{d}q^+\hspace{-0.1cm} \int_0^{\infty}\hspace{-0.2cm}\mathrm{d}q^-\hspace{-0.1cm}\int\mathrm{d}\Omega_{d-2}\left(\frac{2n_{b\bar{b}}}{n_1^+n_1^-} \right)^{-\epsilon}\nn\\
&&\delta(k-q^+-q^-+2\sqrt{q^+q^-}\cos\theta)\frac{n_b\cdot v}{q^+(aq^++q^-)}.
\end{eqnarray}
Introducing two variables $x$ and $y$ such that $q^+=kyx$ and $q^-=ky(1-x)=ky\bar{x}$,
\begin{eqnarray}
S^{(1)}_{bt}(k,\mu)=\frac{g_s^2C_F\mu^{2\epsilon}}{(2\pi)^{d-1}}\left(\frac{2n_{b\bar{b}}}{n_1^+n_1^-} \right)^{-\epsilon} k^{-1-2\epsilon} \int\mathrm{d}\Omega_{d-2}\int_0^{1}\hspace{-0.2cm}\mathrm{d}x
x^{-1-\epsilon}\frac{(1-2\sqrt{x\bar{x}}\cos\theta)^{2\epsilon}\bar{x}^{-\epsilon}}{ax+\bar{x}}.
\end{eqnarray}
The singularity in the integrand can be isolated by
\begin{eqnarray}
x^{-1-\epsilon}=-\frac{1}{\epsilon}\delta(x)+\left(\frac{1}{x}\right)_+-\epsilon\left(\frac{{\rm ln}x}{x}\right)_+ + \mathcal{O}(\epsilon^2).
\end{eqnarray}
After completing the above three parts of the integration separately and expanding
\begin{eqnarray}
\frac{1}{k^+}\left(\frac{\tilde{\mu}}{k^+}\right)^{2\epsilon}=-\frac{1}{2\epsilon}\delta(k^+)+
\left[\frac{1}{k^+}\right]_{\star}^{[k^+,\tilde{\mu}]}-2\epsilon\left[\frac{1}{k^+}{\rm ln}\frac{k^+}{\tilde{\mu}}\right]_{\star}^{[k^+,\tilde{\mu}]}+ \mathcal{O}(\epsilon^2),
\end{eqnarray}
we get the divergent and finite parts
\begin{eqnarray}
S^{(1)}_{bt,div}(k,\mu)&=&\frac{2C_F\alpha_s(4\pi\mu^2 e^{-\gamma_E})^{\epsilon}}{4\pi}
\biggl\{ \frac{\delta(k)}{\epsilon^2}- \frac{2}{\epsilon} \biggl[ \frac{1}{k} \biggr]_{\star}^{[k,\tilde{\mu}]} \biggr\},\\
S^{(1)}_{bt,fin}(k,\mu)&=&\frac{2C_F\alpha_s}{4\pi}\biggl\{ 4\biggl[ \frac{\ln \frac{k}{\tilde{\mu}}}{k} \biggr]_{\star}^{[k,\tilde{\mu}]}+\delta(k)c_{bt}^S \biggr\},
\end{eqnarray}
with
$c_{bt}^S=-{\rm ln}^2(1+\frac{1}{a})-2{\rm Li}_2(\frac{1}{1+a})+\frac{\pi^2}{12}$.

In the same method, we can get
\begin{eqnarray}
S^{(1)}_{tt,div}(k,\mu)&=&\frac{2C_F\alpha_s(4\pi\mu^2 e^{-\gamma_E})^{\epsilon}}{4\pi}
\biggl\{ \frac{\delta(k)}{\epsilon} \biggr\},\\
S^{(1)}_{tt,fin}(k,\mu)&=&\frac{2C_F\alpha_s}{4\pi}\biggl\{ -\biggl[ \frac{2}{k} \biggr]_{\star}^{[k,\tilde{\mu}]}+\delta(k)c_{tt}^S \biggr\},
\end{eqnarray}
with $c_{tt}^S=2{\rm ln}(1+\frac{1}{a})$.

When performing the Laplace transformation from $S(k,\mu)$ to $\widetilde{s}(L,\mu)$, we use the following replacements:
\begin{eqnarray}
  \biggl[ \frac{\ln \frac{k}{\mu}}{k} \biggr]_{\star}^{[k,\mu]} &\to& \frac{L^2}{2}+\frac{\pi^2}{12}, \\
  \biggl[ \frac{1}{k} \biggr]_{\star}^{[k,\mu]}&\to& L.
\end{eqnarray}

\section{anomalous dimensions}\label{sec:anodim}
The various anomalous dimensions
needed in our calculations can be found, e.g., in~\cite{Becher:2006mr,Becher:2007ty,Becher:2009th}. We
list them below for the convenience of the reader. The QCD
$\beta$
function is
\begin{equation}
 \beta (\alpha_s) = -2 \alpha_s \left[ \beta_0
\frac{\alpha_s}{4\pi} + \beta_1 \left(
\frac{\alpha_s}{4\pi} \right)^2 + \cdots \right],
\end{equation}
with expansion coefficients
\begin{eqnarray}
 \beta_0 &=& \frac{11}{3} C_A - \frac{4}{3} T_F n_f,
\nn
\\
\beta_1 &=& \frac{34}{3}C^2_A - \frac{20}{3} C_A T_F n_f -
4C_F T_F n_f,
\nn
\\
\beta_2 &=& \frac{2857}{54} C^3_A + \left( 2 C^2_F -
\frac{205}{9} C_F C_A - \frac{1415}{27} C^2_A \right) T_F
n_f + \left( \frac{44}{9}C_F + \frac{158}{27} C_A \right)
T^2_F n^2_f,
\end{eqnarray}
where $C_A=3$, $C_F=4/3$, $T_F=1/2$ for QCD, and $n_f$ is
the number of active quark flavors.

The cusp anomalous dimension is
\begin{equation}
\label{cuspa}
 \Gamma_{\rm cusp} (\alpha_s) = \Gamma_0\frac{\alpha_s}{4\pi} +
\Gamma_1 \left(
\frac{\alpha_s}{4\pi} \right)^2 + \cdots,
\end{equation}
with
\begin{eqnarray}
 \Gamma_0 &=& 4C_F,
\nn
\\
\Gamma_1 &=& 4C_F \left[ \left( \frac{67}{9} -
\frac{\pi^2}{3} \right) C_A - \frac{20}{9} T_F n_f \right],
\nn
\\
\Gamma_2 &=& 4C_F \left[ C^2_A \left(
\frac{245}{6} -
\frac{134}{27}\pi^2 + \frac{11}{45}\pi^4 +
\frac{22}{3}\zeta_3 \right) + C_A T_F n_f \left(
-\frac{418}{27} + \frac{40}{27}\pi^2 - \frac{56}{3}\zeta_3
\right)
\right.
\nn
\\
&&\left. + C_F T_F n_f \left( -\frac{55}{3} + 16 \zeta_3
\right) - \frac{16}{27} T^2_F n^2_f \right].
\end{eqnarray}

The other anomalous dimensions are expanded as
Eq.~(\ref{cuspa}), and their expansion coefficients are
\begin{eqnarray}
 \gamma^{0}_q &=& -3C_F,
\nn
\\
\gamma^{1}_q &=& C^2_F\left(-\frac{3}{2}+2\pi^2-24
\zeta_3\right) + C_F
C_A \left( -\frac{961}{54}-\frac{11}{6}\pi^2 + 26\zeta_3
\right) + C_F T_F n_f \left( \frac{130}{27} +
\frac{2}{3}\pi^2 \right),
\nn
\\
\gamma^{0}_Q &=& -2C_F,
\nn
\\
\gamma^{1}_Q &=& C_F C_A \left( \frac{2}{3}\pi^2 -
\frac{98}{9} - 4 \zeta_3 \right) + \frac{40}{9} C_F T_F n_f,
\nn
\\
\gamma^{0}_\phi &=& 3C_F,
\nn
\\
\gamma^{1}_\phi &=& C^2_F \left( \frac{3}{2} - 2 \pi^2 +
24 \zeta_3 \right) + C_F C_A \left( \frac{17}{6} +
\frac{22}{9} \pi^2 - 12 \zeta_3 \right) - C_F T_F n_f
\left( \frac{2}{3} + \frac{8}{9} \pi^2 \right),
\nn
\\
\gamma^{0}_j &=& -3C_F,
\nn
\\
\gamma^{1}_j &=& C^2_F \left( -\frac{3}{2} + 2 \pi^2-
24 \zeta_3 \right) + C_F C_A \left( -\frac{1769}{54} -
\frac{11}{9} \pi^2 + 40 \zeta_3 \right)
\nn
\\
&& + C_F T_F n_f
\left( \frac{242}{27} + \frac{4}{9} \pi^2 \right).
\end{eqnarray}
$\gamma^V_{up}$, $\gamma^V_{dn}$ and $\gamma^S$ can be obtained from the
anomalous dimensions above through the following equations:
\begin{eqnarray}
  \gamma^V_{up} &=& 2 \gamma_q ,\nn\\
  \gamma^V_{dn} &=& \gamma_q + \gamma_Q,\nn\\
\gamma^S &=& - \gamma_\phi - \gamma^V_{dn}  .
\label{eqs:anormdim}
\end{eqnarray}

\bibliography{t-channel-top}{}

\end{document}